\documentclass[prb,twocolumn,amsmath,amssymb,floats,showpacs,floatfix]{revtex4-2}
\usepackage{graphicx}
\usepackage{bm}
\usepackage{amsmath}
\usepackage{amsfonts}
\usepackage{amsbsy}
\usepackage{physics}
\usepackage{verbatim}
\usepackage{color}
\usepackage{soul}
\usepackage{float}
\usepackage[nameinlink]{cleveref}
\crefmultiformat{figure}{~#2#1#3}{,\!~#2#1#3}{, #2#1#3}{ and~#2#1#3}
\crefname{figure}{}{}
\usepackage{placeins}

\begin{document}


%
\title{Inductive microwave response of Yu-Shiba-Rusinov states}

\author{C. Hermansen$^{1}$}
\author{A. Levy Yeyati$^{2}$}
\author{J. Paaske$^{1}$}

\affiliation{$^{1}$Center for Quantum Devices, Niels Bohr Institute, University of Copenhagen, 2100 Copenhagen, Denmark}
\affiliation{$^{2}$Departamento de F\'isica Te\'orica de la Materia Condensada, Condensed Matter Physics Center (IFIMAC), and Instituto Nicol\'as Cabrera, Universidad Aut\'onoma de Madrid, 28049 Madrid, Spain}

\date{\today}
\begin{abstract}
We calculate the frequency-dependent admittance of a phase-biased Josephson junction spanning a magnetic impurity or a spinful Coulomb-blockaded quantum dot. The local magnetic moment gives rise to Yu-Shiba-Rusinov bound states, which govern the sub-gap absorption as well as the inductive response. We model the system by a superconducting spin-polarized exchange-cotunnel junction and calculate the linear current response to an AC bias voltage, including its dependence on phase bias as well as particle-hole, and source-drain coupling asymmetry. The corresponding inductive admittance is analyzed and compared to results of a zero-bandwidth, as well as an infinite-gap approximation to the superconducting Anderson model. All three approaches capture the interaction-induced $\boldsymbol{0}-\boldsymbol{\pi}$ transition, which is reflected as a discontinuity in the adiabatic inductive response. 
\end{abstract}

\maketitle

\section{Introduction}

While the understanding of the Josephson effect in terms of Andreev states was achieved long ago~\cite{Kulik1970}, it was not until recently that direct evidence for these states was obtained experimentally from microwave spectroscopy of atomic contacts~\cite{Bretheau2013Jul, Janvier2015Sep}. Microwave circuit-QED (cQED) techniques~\cite{Blais2021May} have also been employed to study the Josephson effect in novel types of weak links comprised by hybrid nanostructures combining semiconducting nanowires with epitaxially deposited superconductors~\cite{deLange2015Sep, vanWoerkom2017Sep, Hays2018Jul, Tosi2019Jan, Hays2020Nov, Kringhoj2020Jun, Metzger2021Jan, Zellekens2021Dec}. Many of the observed features in these experiments have been explained in terms of non-interacting junction models~\cite{Beenakker1991Dec, Furusaki1999May, Ivanov1999Apr, Zazunov2003Feb}, possibly including the effects of spin-orbit coupling~\cite{Park2017Sep, Tosi2019Jan, Hays2020Nov, Metzger2021Jan} and of electrostatic gating~\cite{deLange2015Sep, vanWoerkom2017Sep, Hays2018Jul, Tosi2019Jan, Kringhoj2020Jun}. However, there is also evidence that interactions can play an important role in these devices, especially near pinch off when charge localization becomes important and quantum dot physics can be expected~\cite{Chang2015Mar, Kringhoj2020Jun}.

Recent work by Kurilovich et al.~\cite{Kurilovich2021May} addressed the microwave response of finite-length weak links, taking into account both a finite dwell time on the nanowire as well as weak Coulomb interaction. More specifically, the inverse dwell time was represented by the tunnel broadening, $\Gamma$, within a single-orbital Anderson model with two superconducting leads with gap $\Delta$, and interactions were included perturbatively for $U\ll\Gamma+\Delta$. 

\begin{figure}[b]
\includegraphics[width=0.9\columnwidth]
{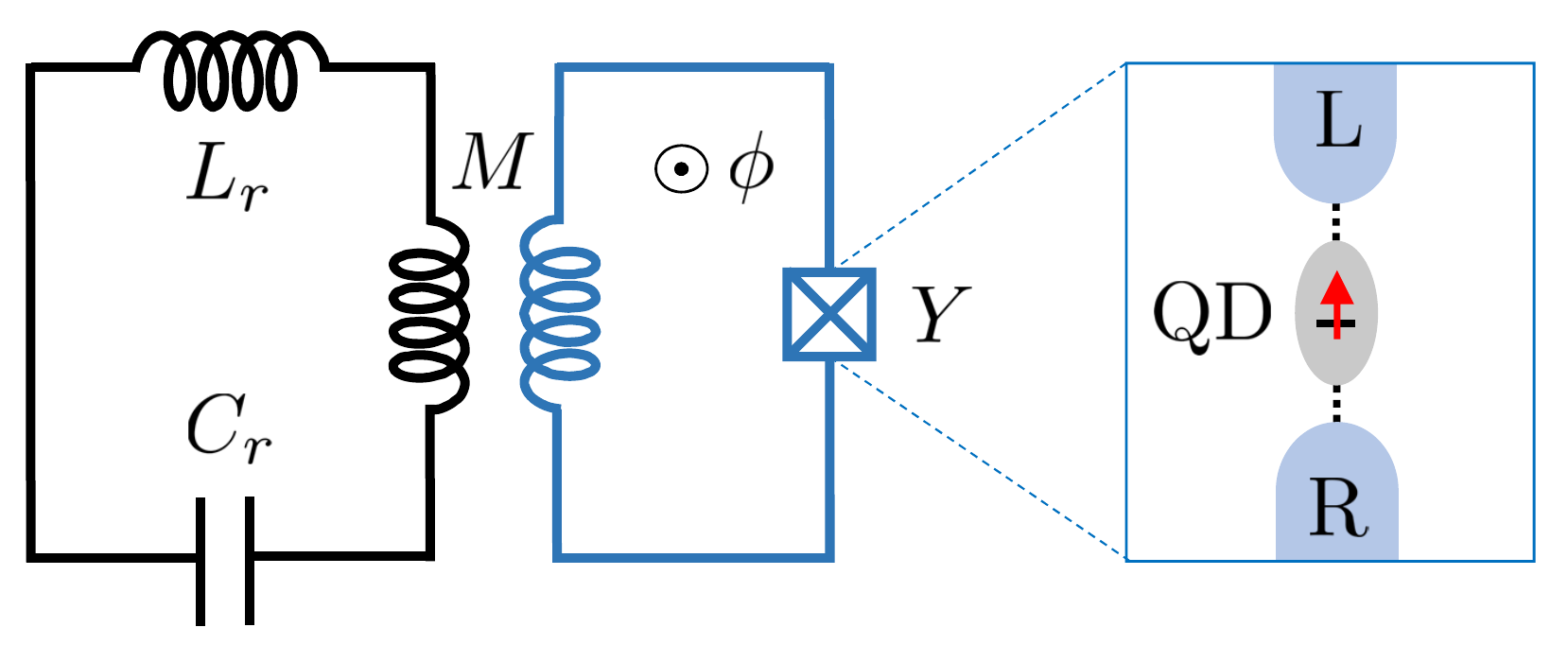}\caption{Illustration of a flux-biased superconducting circuit (blue) spanning a magnetic impurity or a gateable Coulomb-blockaded quantum dot with a single spin-1/2 particle giving rise to YSR sub-gap states. The device is coupled via a weak mutual inductance, $M$, to a microwave resonator (black), chosen here as a simple lumped element with capacitance, $C_r$, and inductance, $L_r$. The resonator induces a weak AC bias voltage across the S-QD-S junction with admittance $Y$. This gives rise to a current response, which in turn shifts and damps the resonator.}
\label{fig:device}
\end{figure}
Here we address the strongly interacting regime, $U\gg\Gamma,\Delta$, describing the weak link as a Coulomb-blockaded quantum dot (QD). Whereas even-occupied dots behave merely as superconducting (co)tunnel junctions~\cite{Kos2013May}, odd-occupied dots provide a local magnetic moment, which gives rise to Yu-Shiba-Rusinov (YSR)~\cite{Yu1965, Shiba1968, Rusinov1969} states inside the gap. We calculate the linear inductive current response of a magnetic impurity or spin-full Coulomb-blockaded QD placed in a superconducting junction coupled to a microwave resonantor by a weak mutual inductance, as illustrated in Fig.~\ref{fig:device}. More specifically, this entails an analysis of the frequency-dependent admittance, $Y(\omega)$, as a function of phase-bias, particle-hole asymmetry, source-drain coupling asymmetry and coupling strength, governing the corresponding frequency shift and damping rate for the resonator, which can be observed with cQED techniques~\cite{Girvin2011, Hays2018Jul, Tosi2019Jan, Hays2020Nov}.

YSR states have been observed in bias spectroscopy of quantum dots based on carbon nanotubes~\cite{Buitelaar2002Dec, Eichler2007Sep, Grove-Rasmussen2009Apr, Pillet2010Dec, Pillet2013Jul, Chang2013May, Kumar2014Feb, Delagrange2016May}, self-assembled InAs quantum dots~\cite{Buizert2007Sep, Deacon2010Feb}, as well as in InAs nanowires~\cite{Lee2014Jan, Chang2015Mar, Jellinggaard2016Aug, Lee2017May, Grove-Rasmussen2018Jun}. YSR states can also support a supercurrent through a Coulomb-blockaded superconductor-quantum-dot-superconductor (S-QD-S) junction, and have been shown to give rise to a gate-voltage induced $\boldsymbol{\pi}-\boldsymbol{0}$ transition~\cite{Cleuziou2006Oct, Jorgensen2007Aug, Maurand2012Feb, Delagrange2015Jun, Delagrange2016May, EstradaSaldana2018Dec, Bouman2020Dec}, which corresponds to a change, in ground state from a spin doublet to a spin singlet in which the dot-spin has been screened by quasiparticles in the leads~\cite{Rozhkov1999Mar, Vecino2003Jul, Martin-Rodero2011Dec, Kirsanskas2015Dec, Meden2019Feb}. This transition is induced by increasing the charge fluctuations on the dot either by gating, or by increasing the tunnel coupling to make $\Gamma/U$ larger. As reviewed in Refs.~\onlinecite{Martin-Rodero2011Dec,Meden2019Feb}, a lot of research has been carried out already regarding sub-gap states and supercurrent through Coulomb-blockaded quantum dots connected to superconducting leads, and the aim of this work is to determine their linear response to microwave radiation. This information is essential to analyze the detection of YSR states in cQED experiments.

At the outset, microwaves must be expected to couple only weakly to a Coulomb-blockaded quantum dot, since only virtual charge fluctuations are possible. Nevertheless, with two leads tunnel coupled to the dot, a high charging energy can be overcome by strong correlation effects, as for the spin-flip induced Kondo correlations giving rise to a perfectly transmitting cotunnel junction in the normal state for temperatures smaller than the Kondo temperature, $T\ll T_{K}$~\cite{Glazman1988, vanderWiel2000Sep}. When the leads become superconducting, a similar screening takes place, in terms of a sub-gap level crossing between the unscreened doublet state and the screened singlet, which occurs near $T_{K}\sim\Delta$~\cite{Satori1992Sep, Siano2004Jul, Choi2004Jul, Bauer2007, Moca2021Oct}. For $T_{K}\ll\Delta$, the cotunnel junction hosts sub-gap states very close to the gap, and the corresponding $\boldsymbol{\pi}$ junction has a very low critical current of the order of the exchange-cotunnel coupling squared, $I_c\approx (\nu_{F}J)^{2}e\Delta/\hbar$~\cite{Kirsanskas2015Dec}. For $T_{K}\gg\Delta$, however, the nearly perfectly transmitting $\boldsymbol{0}$ junction has $I_c\approx e\Delta/\hbar$~\cite{Glazman1989} with sub-gap states close to the gap at $\varphi=0$ and close to zero energy at $\varphi=\pi$. Overall, the Coulomb-blockaded quantum dot acts as a tunable cotunnel junction characterized by its normal-state transmission, which reaches one for symmetric couplings and $T_{K}\gg\Delta, T$. If the QD is either even-occupied (spinless) or odd-occupied (spinful) at very strong coupling,  $T_{K}\gg\Delta, T$, one may therefore expect a microwave response much like that of the superconducting weak link studied in Ref.~\onlinecite{Kos2013May}. In general, an odd-occupied QD can be tuned from $\boldsymbol{\pi}$, to $\boldsymbol{0}$ junction behavior by increasing $T_{K}/\Delta$, passing through $\boldsymbol{\pi^\prime}$, and $\boldsymbol{0^\prime}$ behavior~\cite{Rozhkov1999Mar, Kirsanskas2015Dec, Delagrange2016May} with discontinuous current-phase relations. These regimes of different current-phase relations will support different inductive responses from the flux-derivative of the supercurrent, alongside with the corresponding current-carrying transitions between phase-dependent sub-gap YSR states, which add up to a variety of different possible microwave responses.


The paper is organized as follows: In Sec.~\ref{sec:AM} we define the S-QD-S junction as an Anderson model with superconducting leads and implement the coupling to the resonator through a time-dependent phase-difference. 

In Sec.~\ref{sec:lrac}, we work out the linear response relations and relate the frequency shift and damping rate of the resonator to the admittance of the junction. 

Sec.~\ref{sec:IG} is devoted to an analytical solution for the infinite-gap limit. This result captures the dynamics of the proximity effect, and reproduces the $\Delta\gg U$-limit of the current-current response function found in Ref.~\onlinecite{Kurilovich2021May}. It also illustrates the vanishing of the current matrix element for microwave absorption when the sub-gap transition energy becomes smaller than $U/2$, and the system switches from an even-parity singlet, to an odd-parity doublet ground state, corresponding to a $\boldsymbol{\pi}$ junction. 

In Sec.~\ref{sec:chYSR} we address the response of YSR states calculated within a polarized-spin approximation, which has been used earlier to determine the phase dispersion of the YSR states~\cite{Kirsanskas2015Dec}. This approach is based on a Kondo model, derived by means of a Schrieffer-Wolff transformation of the single-orbital Anderson model, which eliminates charge fluctuations and leaves only inter-lead charge transport. This tacitly implies zero dwell time on the dot and hence a short-junction description for which the current response must vanish when decoupling one of the leads. Whereas this is in stark contrast to the infinite-gap limit, which retains the charge fluctuations and hence currents between the dot and the individual leads, both models share the presence of a doublet phase for large enough $U$. 

In Sec.~\ref{sec:ZBW} we demonstrate that the numerically tractable zero-bandwidth model~\cite{Allub1981Feb, Grove-Rasmussen2018Jun, EstradaSaldana2018Dec} qualitatively interpolates between the results of the infinite-gap model and the polarized-spin approximation, capturing the salient features of resonant microwave response for both $\Delta\gg U, \Gamma$ and $U\gg\Delta,\Gamma$. Finally, we discuss the response of excited states, which will be of relevance to actual experiments with finite parity-decay times due to quasiparticle poisoning.


\section{Anderson model with superconducting leads}\label{sec:AM}

We model the S-QD-S Josephson junction by a single-orbital Anderson model, $H=H_{LR}+H_D+H_T$, with BCS mean-field Hamiltonians for the two leads,
\begin{align}
H_{LR}=&\sum_{\alpha k\sigma}\xi_{k\sigma}c^\dagger_{\alpha k \sigma}c_{\alpha k \sigma}\label{am1}\\
&\hspace*{10mm}-\sum_{\alpha k}\left( \Delta_\alpha c^\dagger_{\alpha k \uparrow}c^\dagger_{\alpha -k \downarrow}+\Delta^*_\alpha c_{\alpha -k\downarrow} c_{\alpha k\uparrow}\right),\nonumber
\end{align}
together with a single-orbital quantum dot with energy $\epsilon_{d}$ and charging energy $U$,
\begin{align}
H_D=\sum_\sigma \epsilon_d d^\dagger_\sigma d_\sigma+U n_{d\uparrow} n_{d\downarrow},\label{am2}
\end{align}
and dot-lead tunnelling terms,
\begin{align}
H_T=\sum_{\alpha k \sigma}(t_\alpha c^\dagger_{\alpha k \sigma}d_\sigma+t^*_\alpha d^\dagger_\sigma c_{\alpha k \sigma}).\label{am3}
\end{align}
Left, and right leads are indexed by $\alpha=L,R$, and operators $c^{\dagger}_{\alpha k\sigma}$ and $d^{\dagger}_{\sigma}$ create electrons with spin $\sigma$ in respectively the superconducting lead $\alpha$ with momentum $k$, and the single QD level with energy $\epsilon_{d}$, which can be tuned by a gate voltage. The dot-lead tunnelling amplitudes, $t_{\alpha}$, are assumed to be real and momentum independent, and the two leads are assumed to have BCS gaps of same magnitude and with a phase difference, $\varphi$, i.e. $\Delta_R=\Delta e^{i\varphi/2}$ and $\Delta_L=\Delta e^{-i\varphi/2}$ with $\Delta=|\Delta|$. 

As depicted in Fig.~\ref{fig:device}, we wish to investigate the linear current response of this system to a weak time-dependent magnetic flux, $\delta\phi(t)=\phi_{0}\delta\varphi(t)$, implemented in the Hamiltonian above as a modulation of the phase difference, in terms of the reduced flux quantum, $\phi_{0}=\hbar/2e$. According to the Josephson relation, this comes with an AC bias voltage, $\delta V(t)=\delta\dot{\phi}(t)$, contributing to the Hamiltonian with the term $(\hat{N}_{R}-\hat{N}_{L})e\delta V(t)/2$. These terms are most conveniently dealt with by a time-dependent gauge transformation, which removes the bias voltage from the Hamiltonian and transfers the phase from the order parameters to the tunnel couplings, i.e. $t_{L/R}\rightarrow t_{L/R}e^{\pm i\varphi(t)/4}$, with $\varphi(t)=\varphi+\delta\varphi(t)$. 



\section{Linear response}\label{sec:lrac}
The applied AC perturbation now enters the Hamiltonian only via the time-dependent phase shift $\delta\varphi(t)=\delta\phi(t)/\phi_{0}$, and results in the following perturbation:
\begin{align}
\delta\hat{H}(t)=\hat{I}\delta\phi(t),
\end{align}
with current operator $\hat{I}=\partial_{\phi}\hat{H}$. In a given state, $\ket{n}$, be it an excited state, or the ground state of the unperturbed system, a finite $DC$ flux bias induces a supercurrent, $I_{n}=\langle\hat{I}\rangle_{n}$. A slight change in flux, $\delta\phi$, gives rise to an additional linear-response current
\begin{align}
\delta\langle\hat{I}(t)\rangle_{n}=\langle\delta\hat{I}(t)\rangle_{n}+\int^\infty_{-\infty}\!\!\!dt'\,\chi_{n}^{R}(t-t')\delta\phi(t'),
\label{linres_current}
\end{align}
expressed in terms of the retarded current-current response function
\begin{align}
\chi_{n}^{R}(t-t')=-\frac{i}{\hbar}\theta(t-t')\langle[\hat{I}(t),\hat{I}(t')]\rangle_{n},
\label{response_function}
\end{align}
and the first order change in the current operator itself:
\begin{align}
\langle\delta\hat{I}(t)\rangle_{n}=\langle\partial_{\phi}\hat{I}\rangle_{n}\delta\phi(t).    
\end{align}
For convenience, we shall use the same symbols in the Fourier-transformed response relation,
\begin{align}\label{eq:LRomega}
\delta\langle\hat{I}(\omega)\rangle_{n}=
\left(\langle\partial_{\phi}\hat{I}\rangle_{n}+\chi_{n}^{R}(\omega)\right)\delta\phi(\omega),
\end{align}
As argued in Ref.~\onlinecite{Kurilovich2021May}, the static response will correspond to an adiabatic change in the supercurrent, i.e. $\delta\langle\hat{I}\rangle_{n}=\partial_{\phi}\langle\hat{I}\rangle_{n}\delta\phi$. Using Eq.~\eqref{eq:LRomega} at zero frequency, this implies that $\langle\partial_{\phi}\hat{I}\rangle_{n}+\chi_{n}^{R}(0)=\partial_{\phi}\langle\hat{I}\rangle_{n}$, whereby
\begin{align}\label{eq:dI}
\delta\langle\hat{I}(\omega)\rangle_{n}=\left(\partial_{\phi}\langle\hat{I}\rangle_{n}
+\delta\chi_{n}(\omega)\right)\delta\phi(\omega),
\end{align}
with $\delta\chi_{n}(\omega)=\chi^{R}_{n}(\omega)-\chi^{R}_{n}(0)$, which vanishes at zero frequency. The linear current response of the system in state $\ket{n}$ may therefore be written simply as \begin{align}
\delta\langle\hat{I}(\omega)\rangle_{n}=\chi_{II,n}(\omega)\delta\phi(\omega)
\end{align}
with the current-current response function
\begin{align}\label{eq:respf}
\chi_{II,n}(\omega)=\partial_{\phi}\langle\hat{I}\rangle_{n}+\delta\chi_{n}(\omega),
\end{align}
corresponding to a junction admittance of
\begin{align}\label{eq:admit}
Y_{n}(\omega)=\frac{i}{\omega}\chi_{II,n}(\omega).
\end{align}
Since the current as well as the zero-frequency part of the response function are purely real, the imaginary parts, $\chi''_{II}(\omega)=\delta\chi''(\omega)={\chi^R}''(\omega)$, can be obtained directly from $\chi^R(\omega)$.

Coupling this circuit to a (lumped-element) resonator, with inductance $L_{r}$ and capacitance $C_{r}$, by a weak mutual inductance, $M\ll L_{r}$, as illustrated in Fig.~\ref{fig:device}, the Fourier transformed equations of motion for the resonator flux, $\phi_{r}$, and charge, $q_{r}$, are modified to
\begin{align}
i\omega \phi_{r}(\omega)=&\,-\frac{q_{r}(\omega)}{C_{r}}\\
i\omega q_{r}(\omega)=&\frac{\phi_{r}(\omega)}{L_{r}}-\chi_{II,n}(\omega)(M/L_{r})^{2}\phi_{r}(\omega).\,
\end{align}
To second order in $M/L_{r}$, the resonator frequency is shifted from $\omega_{r}=1/\sqrt{L_{r}C_{r}}$ to $\omega_{r}-\delta\omega_{r}$, with
\begin{align}\label{eq:shift}
\delta\omega_{r}^{(n)}=\lambda^{2}\frac{\phi_{0}^{2}}{\hbar}\chi'_{II,n}(\omega_{r}),
\end{align}
and damped with rate
\begin{align}\label{eq:damp}
\gamma_{r}=-\lambda^{2}\frac{\phi_{0}^{2}}{\hbar}\chi''_{II,n}(\omega_{r})
\end{align}
in terms of imaginary and real parts of the response function, $\chi_{II,n}=\chi'_{II,n}+i\chi''_{II,n}$. We use this notation for real, and imaginary parts only for the admittance and for the current-current response function. Here the dimensionless coupling constant is given by:
\begin{align}
\lambda=\frac{M}{L_{r}}\sqrt{\frac{\pi Z_{0}}{R_{Q}}},
\end{align}
where $Z_{0}=\sqrt{L_{r}/C_{r}}$ denotes the resonator impedance and $R_{Q}=h/4e^{2}$ the resistance quantum. As demonstrated in Ref.~\onlinecite{Park2020Aug} the corresponding analysis for a quantum mechanical resonator gives the same frequency shift. According to Refs.~\cite{Metzger2021Jan, Hays2021Jul}, the value of $\lambda$ in actual devices typically ranges from $0.05$ to $0.001$.

\section{Infinite-gap limit}\label{sec:IG}

Before studying the response of the S-QD-S junction in the effective cotunnelling model, it is instructive to consider the infinite-gap limit in which $\Delta$ is much larger than all other relevant energy scales. In this limit, the continuum states are eliminated completely and the linear response is readily calculated analytically, and shown to match the corresponding limit of the results in Ref.~\onlinecite{Kurilovich2021May}.

\subsection{Effective proximitized-dot Hamiltonian}

Following Refs.~\onlinecite{Tanaka2007May, Meng2009}, the dot-electron Nambu self-energy due to tunnelling simply reads
\begin{align}
\Sigma_{d}(i\omega_n)=\sum_{\alpha=L,R}t_{\alpha}^{2}\tau^{z}\mathcal{G}_{0,\alpha}(i\omega_n)\tau^{z},
\end{align}
where $\tau^{i}$ denotes the Pauli matrices and with Matsubara lead-electron Nambu Green function
\begin{align} 
\mathcal{G}_{0,\alpha}(i\omega_n)=\frac{\pi\nu_F}{\sqrt{|\Delta|^2-(i\omega_n)^2}}
\begin{pmatrix}
-i\omega_n &&  \Delta_{\alpha}\\ 
\Delta_{\alpha}^{\ast} && -i\omega_n
\end{pmatrix},
\end{align}
defined in terms of a constant density of states near the Fermi level, $\nu_{F}$, assumed without loss of generality to be the same in the two leads. For $\Delta \gg \omega$ and with $\Gamma_{\alpha}=\pi\nu_{F}t_{\alpha}^2$, this simplifies to the constant self-energy 
\begin{equation}
\Sigma_d=
-\begin{pmatrix}
0 && \Gamma_{L}e^{-i\varphi/2}+\Gamma_{R}e^{i\varphi/2} \\ \Gamma_{L}e^{i\varphi/2}+\Gamma_{R}e^{-i\varphi/2} && 0
\end{pmatrix}.
\end{equation}
Integrating out the leads, we therefore arrive at the low-energy effective Hamiltonian
\begin{align}
\!\!\hat{H}_{\infty}=&\sum_{\sigma=\uparrow,\downarrow}\!\!\xi_d d^\dagger_\sigma d_\sigma-\gamma d^\dagger_\uparrow d^\dagger_\downarrow -\gamma^{\ast}d_\downarrow d_\uparrow+\frac{U}{2}(n_d-1)^2,
\end{align}
with $\xi_d=\epsilon_d+U/2$, $n_d=\sum_{\sigma} d^\dagger_\sigma d_\sigma$, and a tunnelling-induced $d$-electron gap given by
\begin{align}
\gamma=\Gamma\left(\sin^2(\theta)e^{i\varphi/2}+\cos^2(\theta)e^{-i\varphi/2}\right) ,  
\end{align}
or on polar form, $\gamma=|\gamma|e^{i\zeta}$, with
\begin{align}
|\gamma|=&\Gamma\sqrt{\cos^{2}(\varphi/2)+\cos^{2}(2\theta)\sin^{2}(\varphi/2)},\\
\zeta=&\arg(\cos(\varphi/2)-i\cos(2\theta)\sin(\varphi/2)).
\end{align}
The left-right coupling asymmetry is parameterized as $(t_L,t_R)/t=(\cos(\theta),\sin(\theta))$ with $t=\sqrt{t_{L}^2+t_{R}^2}$ and $\theta\in[0,\pi/2]$, and tunnelling rate, $\Gamma=\pi\nu_{F}t^2$. 

This many-body Hamiltonian for the proximitized quantum dot has two odd-parity eigenstates, $\ket{\sigma}$, with $\sigma=\uparrow,\downarrow$, and eigenenergy $E_{\sigma}=\xi_{d}$, as well as two even-parity eigenstates,  $\ket{\mp}$, with eigenenergies $E_\mp=\xi_d+U/2\mp E_{A}$, where $E_{A}=\sqrt{\xi_d^2+|\gamma|^2}$. The even-parity states are given by $\ket{-}=u\ket{0}+v e^{i\zeta}\ket{\uparrow\downarrow}$ and $\ket{+}=u\ket{\uparrow\downarrow}-v e^{-i\zeta}\ket{0}$, where $\ket{\uparrow\downarrow}=d^\dagger_\uparrow d^\dagger_\downarrow\ket{0}$, and
\begin{align}
u=\frac{1}{\sqrt{2}}\sqrt{1+\frac{\xi_d}{E_{A}}}\quad{\rm and}\quad v=\frac{1}{\sqrt{2}}\sqrt{1-\frac{\xi_d}{E_{A}}}.
\end{align}

\subsection{Linear response}

The current operator takes the form
\begin{align}
\hat{I}=\partial_{\phi}\hat{H}_{\infty} =-\gamma' d^\dagger_\uparrow d^\dagger_\downarrow-\gamma'^{\ast}d_\downarrow d_\uparrow, 
\end{align}
where $\gamma'=\partial_{\phi}\gamma$. Among the eigenstates ($n=+,-,\uparrow,\downarrow$), only the even-parity states, $|\mp\rangle$, have non-zero matrix elements of the current operator, arising from the following basic elements
\begin{align}
\bra{\mp}d^\dagger_\uparrow d^\dagger_\downarrow\ket{\mp}=&\pm uv e^{-i\zeta},\\
\bra{+}d^\dagger_\uparrow d^\dagger_\downarrow\ket{-}=&u^{2},\\
\bra{+}d_\downarrow d_\uparrow\ket{-}=&-v^{2}e^{2i\zeta}.
\end{align}
Making use of the relation $E_{A}\partial_{\phi}E_{A}=|\gamma|\partial_{\phi}|\gamma|$, this implies the following expectation values of the current
\begin{align}
\langle\hat{I}\rangle_{\mp}=\mp\partial_{\phi}E_{A}
=\pm\frac{\Gamma^{2}\sin^{2}(2\theta)}{\phi_{0}E_{A}}\sin(\varphi).
\end{align}
Using that $(4|\gamma'||\gamma|/\Gamma^{2})^{2}=\sin^{2}(\varphi)\sin^{4}(2\theta)+4\cos^{2}(2\theta)$, the expectation values of the derivative of the current operator becomes
\begin{align}\label{Ideriv}
\langle\partial_{\phi}\hat{I}\rangle_{\mp}=
\mp\left[\partial^{2}_{\phi}E_{A}-\frac{1}{E_{A}}|I_{+-}|^{2}\right],
\end{align}
where $|I_{+-}|^{2}=|\!\bra{+}\hat{I}\ket{-}\!|^{2}
=|\gamma'|^{2}-(\partial_{\phi}E_{A})^{2}$, which may be rewritten as
\begin{align}
|I_{+-}|^{2}=\phi_{0}^{-2}|\gamma|^{-2}\left(\xi_{d}^{2}(\partial_{\varphi}E_{A})^{2}+\frac{1}{4}\Gamma^{4}\cos^{2}(2\theta)\right).
\end{align}
Finally, the Fourier-transformed response function is obtained as
\begin{align}
\chi_{-}^{R}(\omega)=-\chi_{+}^{R}(\omega)=\frac{4E_{A}|I_{+-}|^{2}}{(\hbar\omega+i0_{+})^{2}-4E_{A}^{2}},
\end{align}
the zero-frequency component of which is recognized as the last term in Eq.~\eqref{Ideriv}. Defining $\delta\chi(\omega)=\chi_{-}^{R}(\omega)-\chi_{-}^{R}(0)$, one has
\begin{align}
\delta\chi(\omega)
=&\frac{|I_{+-}|^{2}}{E_{A}}\frac{(\hbar\omega)^{2}}{(\hbar\omega+i0_{+})^{2}-4E_{A}^{2}},
\end{align}
and the total current-current response function may therefore be written simply as
\begin{align}\label{eq:chiphi}
\chi_{\mp,II}(\omega)=&\mp\left(\partial^{2}_{\phi}E_{A}-\delta\chi(\omega)\right).
\end{align}
This result corresponds to the limit $\max(U,\Gamma)/\Delta\rightarrow 0$ of the result obtained recently by Kurilovich et al.~\cite{Kurilovich2021May}. In this limit, of course, the continuum states are never reached and the odd-parity states, $\ket{\sigma}$, therefore have no phase dependence and hence a vanishing admittance. 

\begin{figure}[t]
\includegraphics[width=\linewidth]{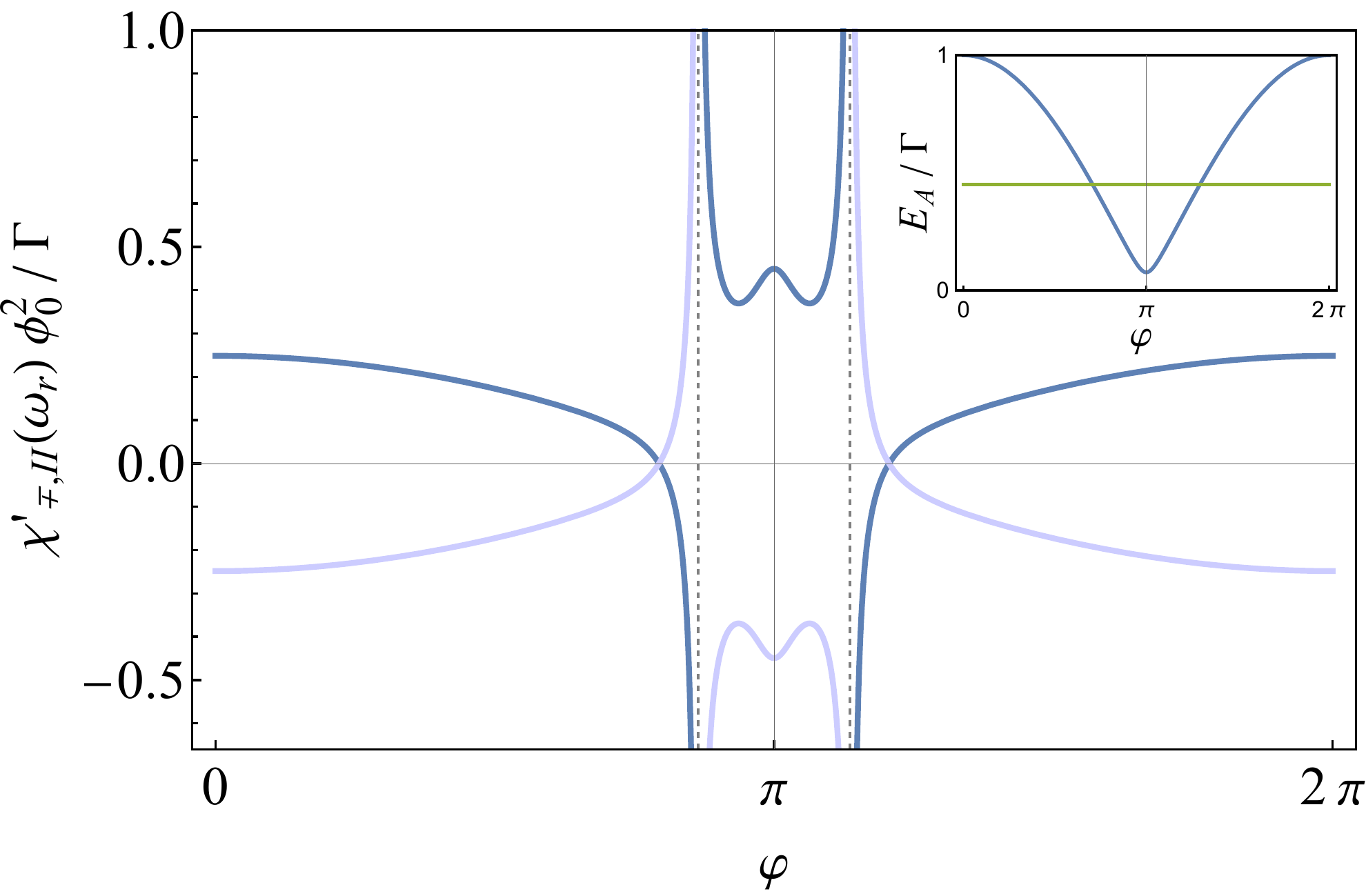}
\caption{Real part of the current-current response function calculated in the infinite-gap limit as a function of phase-difference. Blue (light blue) curve corresponds to $n=-$ ($n=+$). For direct comparison, parameters are chosen as in Fig.4(a) in Ref.~\onlinecite{Kurilovich2021May} to $\omega_{r}=0.45\Gamma$, $\Gamma_{L}=0.30$, $\Gamma_{R}=0.35$, corresponding to $\theta=\pi/3.81$. Gridlines are placed at the values of $\varphi$ at which $2E_{A}=\omega_{r}$. Inset shows $E_{A}$ vs. $\varphi$ (blue) with $\omega_{r}$ indicated by the green line.}
\label{fig:chiphi}
\end{figure}

According to Eq.~\eqref{eq:shift}, the shift of the resonance frequency, $\delta\omega_{r}$, can be obtained from the real part of the response function. This is plotted in Fig.~\ref{fig:chiphi} at a frequency which matches the two-quasiparticle excitation energy for two values of $\varphi$ close to $\pi$, marked by the gridlines. At these two values of $\varphi$, the resonator experiences a formally divergent frequency shift.

\begin{figure*}[t]
\centering
\includegraphics[width=0.45\linewidth]{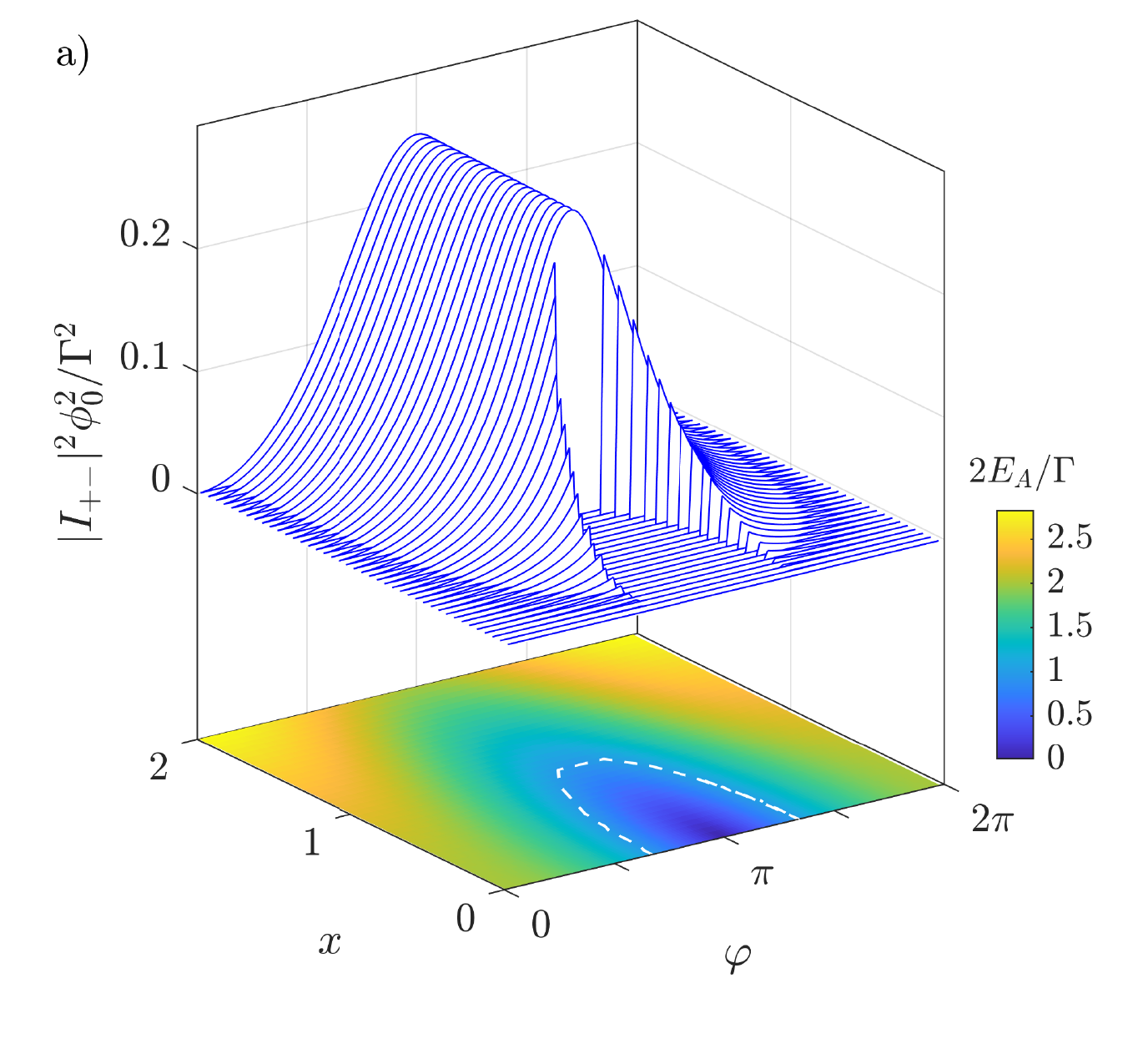}\quad
\includegraphics[width=0.45\linewidth]{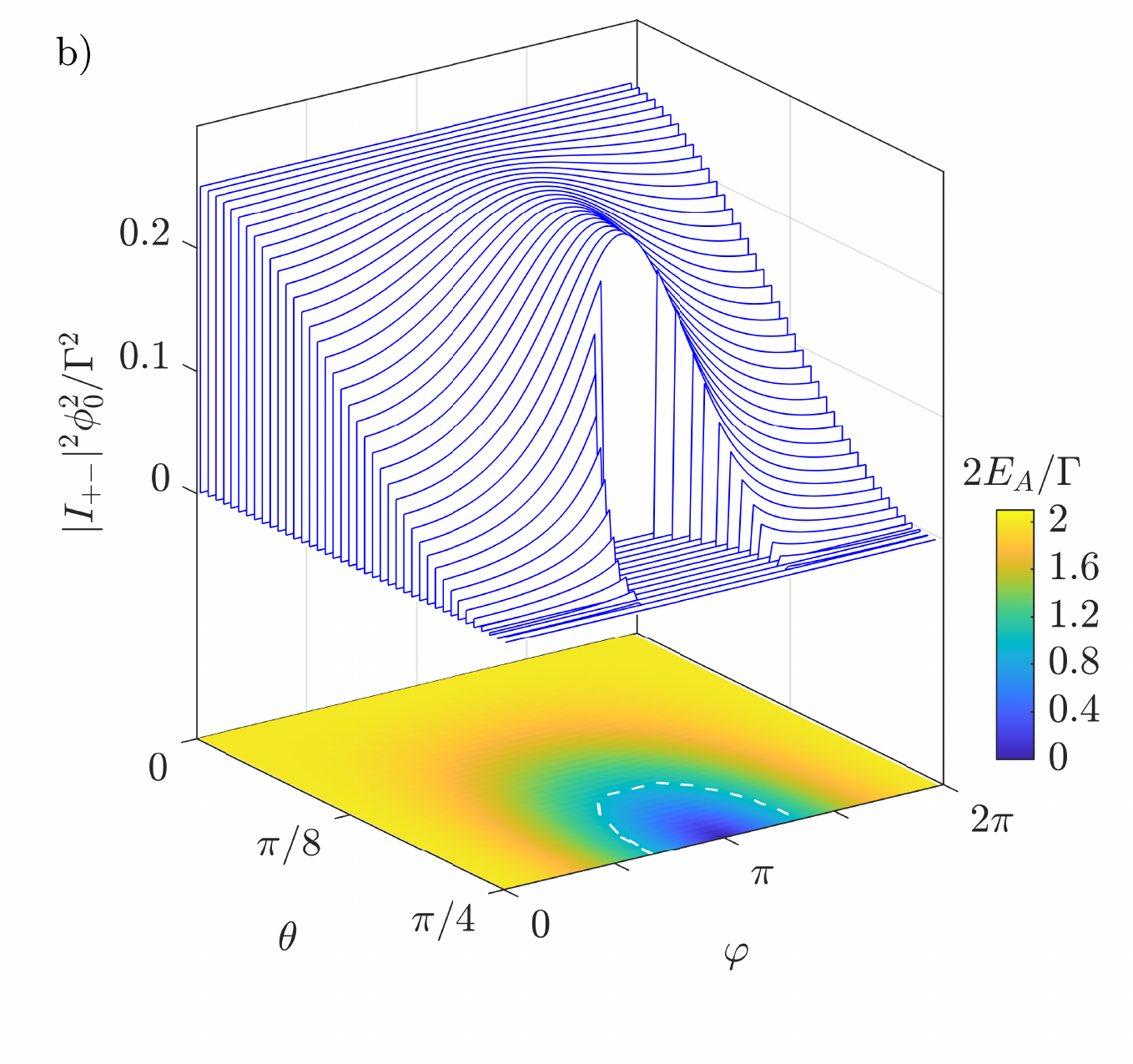}
\caption{Current matrix element determining $\chi_{\pm,II}''$ at resonance (blue curves) calculated in the infinite-gap limit together with the corresponding transition energy, $\hbar\omega=2E_{A}$, (bottom) plotted as a function of phase bias and, respectively, (a) particle-hole asymmetry, $x=2\xi_{d}/U$, at symmetric coupling $\theta=\pi/4$, i.e. $\Gamma_{L}=\Gamma_{R}$ in or (b) coupling asymmetry, $\theta$, at particle-hole symmetry $x=0$, i.e. $\epsilon_{d}=-U/2$ in b). In both plots the coupling strength is $\Gamma/U=1$. White dashed lines delimit the region of odd-parity (doublet) ground state for which the matrix element for microwave absorption is zero. Inside this doublet region the excitation energy refers to the transition between the two excited singlets, $|-\rangle$ and $|+\rangle$.}
\label{fig:infgaptheta}
\end{figure*}

The imaginary part of the response function, corresponding to the rate of energy transfer between the resonator and the superconducting circuit from Eq.~\eqref{eq:damp}, can be found as 
\begin{align}
\chi''_{\mp,II}(\omega)&=\pm\delta\chi''(\omega)\\
&=\mp\pi|I_{+-}|^{2}\left[\delta(\hbar\omega-2E_{A})
-\delta(\hbar\omega+2E_{A})\right]\nonumber,
\end{align}
with weights of the $\delta$-functions given by the current matrix element. These weights are plotted together with the resonance frequency in Fig.~\ref{fig:infgaptheta} as a function of phase difference, particle-hole, and left-right, asymmetry. In these plots, we have chosen $U$ to be large enough to support a region in which the doublet state, $\ket{\sigma}$, is the ground state. In this region, where $U>2E_{A}$, there is no current matrix element for microwave absorption and therefore no current response in the infinite-gap limit. 

\section{Finite gap and YSR states}\label{sec:chYSR}

\subsection{Effective cotunnelling model}

In the strongly interacting limit, $U\gg\Delta,\Gamma$, the Coulomb-blockaded quantum dot has a definite charge which is controlled by a gate voltage. In the case of odd occupancy, the dot comprises a spin-1/2 magnetic impurity, which gives rise to Kondo effect when the leads are normal, and Yu-Shiba-Rusinov (YSR) states in the case of superconducting leads. As with the Kondo effect, the problem may be simplified by eliminating charge fluctuations via a Schrieffer-Wolff transformation of the original Hamiltonian in Eqs.~(\ref{am1}-\ref{am3}). Following Ref.~\onlinecite{Kirsanskas2015Dec}, one arrives at the effective exchange cotunnelling Hamiltonian 
\begin{equation}\label{eq:ysrham}
H=H_{LR}+H_J+H_W
\end{equation}
with exchange, and potential cotunnelling terms
\begin{align}
H_J=&\sum_{\alpha k \sigma, \alpha' k'}J_{\alpha\alpha'}\vec{S}\cdot\vec{\sigma}c^\dagger_{\alpha k\sigma} c_{\alpha' k'\sigma},\\
H_W=&\sum_{\alpha k \sigma, \alpha' k'}W_{\alpha\alpha'} c^\dagger_{\alpha k\sigma} c_{\alpha' k'\sigma},
\end{align}
written with the vector of Pauli matrices, $\vec{\sigma}$, and valid for weak coupling to the leads and odd occupancy of the dot, i.e. $\Gamma_{L,R}\ll-\epsilon_{d}, \epsilon_{d}+U$. The cotunneling amplitudes are matrices in $(L/R)$ lead space, given by 
$J_{\alpha\alpha'}=J\Theta_{\alpha\alpha'}$, and $W_{\alpha\alpha'}=W\Theta_{\alpha\alpha'}$, where
\begin{align}\label{eq:gw}
\Theta_{\alpha\alpha'}=&\begin{pmatrix}
\cos^2(\theta) && \sin(\theta)\cos(\theta)e^{i\varphi/2}\\
\sin(\theta)\cos(\theta)e^{-i\varphi/2} && \sin^2(\theta)
\end{pmatrix}
\end{align}
with dimensionless couplings, $g=\pi\nu_{F}J$ and $w=\pi\nu_{F}W$, given by
\begin{align}
w=\frac{\Gamma}{U}\frac{2x}{1-x^2}, \quad{\rm and}\quad g=\frac{\Gamma}{U}\frac{4}{1-x^2},
\label{parametrization}
\end{align}
where $x=1+2\epsilon_{d}/U$ measures the departure from the particle-hole symmetric point ($x=0$). Even though the matrix~\eqref{eq:gw} has a zero eigenvalue, indicating that only a single scattering channel is involved, at finite phase bias ($\varphi\neq 0$) the anomalous terms in the BCS leads will mix these scattering channels and render this a genuine two-channel scattering problem, except for the case $\varphi=0$~\cite{Kirsanskas2015Dec}. As we shall see below, this channel index is reflected in the sub-gap states, and will play a crucial role for the absorption of microwaves. 

The full spin-rotation-invariant exchange cotunnelling model remains analytically intractable. Depending on the ratio of the Kondo temperature to the BCS gap, $T_{K}/\Delta$, a gradual quasiparticle screening of the impurity (dot) spin takes place much like in the Kondo problem, and this can be captured exactly by Numerical Renormalization Group (NRG) calculations~\cite{Satori1992Sep, Bauer2007, Pillet2013Jul, Moca2021Oct}. In the weak-coupling limit, $g\ll 1$, the model is amenable to a  zero-bandwidth (ZBW) approximation in which the leads are represented merely by a single quasiparticle spin doublet of energy $\Delta$~\cite{Kirsanskas2015Dec}, and even at stronger couplings, this approximation is known to reproduce many features of the NRG solution correctly, up to a rescaling of $\Gamma$~\cite{Grove-Rasmussen2018Jun, EstradaSaldana2018Dec} and will be considered in section~\ref{sec:ZBW} below. 

In this section, we shall restrict our attention to the polarized-spin approximation, in which the quantum dot spin is treated as a classical variable with a fixed direction, $\vec{S}\approx S\hat{z}$. Within this approximation, the problem corresponds to a phase-biased two-channel version of the original problem considered by Yu, Shiba, and Rusinov~\cite{Yu1965,Shiba1968,Rusinov1969}.

\subsection{Nambu Green function}
Within the polarized-spin approximation, the Hamiltonian~\eqref{eq:ysrham} can be expressed as
\begin{equation}\label{Heff}
H=\frac{1}{2}\sum_{\alpha\alpha'\!, kk'\!,\sigma}\psi^\dagger_{\alpha k\sigma}h_{\alpha\alpha'\!,kk'\!,\sigma}\psi_{\alpha' k'\sigma},
\end{equation}
in terms of spin-dependent Nambu spinors, $\psi_{\alpha k\sigma}^\dagger=(c_{\alpha k\sigma}^\dagger, c_{\alpha -k\bar{\sigma}})$, satisfying the anomalous commutation relations
\begin{align}
\{\psi_{\alpha k\sigma}^\dagger, \psi_{\beta k' \sigma'}\}=&\,\delta_{\alpha,\beta}\delta_{k,k'}\delta_{\sigma,\sigma'}\tau^{0}\\
\{\psi_{\alpha k\sigma}, \psi_{\beta k' \sigma'}\}=&\,\delta_{\alpha,\beta}\delta_{k,-k'}\delta_{\sigma,\bar{\sigma}'}\tau^{x},\nonumber
\end{align}
and where $h_{\alpha\alpha'\!, kk'\!,\sigma}=(\xi_{k}\tau^{3}-\sigma\Delta\tau^{1})\delta_{\alpha,\alpha'}\delta_{k,k'}+V_{\alpha\alpha',\sigma}$, with cotunnelling amplitudes 
\begin{align}
V_{\alpha\alpha',\sigma}=
\begin{pmatrix}
V^{\mathcal R}_{L\sigma} & V^{\mathcal T}_{\sigma}\\
\big(V^{\mathcal T}_{\sigma}\big)^{\dagger} & V^{\mathcal R}_{R\sigma}
\end{pmatrix}_{\alpha\alpha'}
\end{align}
with $\alpha,\alpha'\in\{L,R\}$. Here, reflection, and transmission terms have the following Nambu matrix elements
\begin{align}
V^{\mathcal R}_{L\sigma,ij}=&\,(\sigma JS\tau_{ij}^{0}+W\tau_{ij}^{3})\cos^2(\theta),\\
V^{\mathcal R}_{R\sigma,ij}=&\,(\sigma JS\tau_{ij}^{0}+W\tau_{ij}^{3})\sin^2(\theta),\\
V^{\mathcal T}_{\sigma,ij}=&\, (\sigma JS\tau_{ij}^{0}+W\tau_{ij}^{3}) e^{\tau_{ij}^{3}i\varphi/2}\cos(\theta)\sin(\theta).
\end{align}
The momentum-summed Matsubara Green functions,
\begin{align}
\mathcal{G}_{\alpha\alpha',\sigma}(\tau-\tau')=-\sum_{kk'}\langle T_\tau \psi_{\alpha k\sigma}(\tau)\psi_{\alpha' k'\sigma}^\dagger(\tau')\rangle,
\end{align}
are found by solving the Dyson equation
\begin{align}
\mathcal{G}_{\alpha\alpha',\sigma}(i\omega_n)=&\\
&\hspace*{-14mm}\tilde{\mathcal{G}}^{(0)}_{\alpha,\sigma}(i\omega_n)\left(\delta_{\alpha\alpha'}
+\sum_{\alpha''}\Sigma_{\alpha\alpha'',\sigma}\mathcal{G}_{\alpha''\alpha',\sigma}(i\omega_n)\right),\nonumber
\end{align}
where $\tilde{\mathcal{G}}^{(0)}_{\alpha\sigma}(i\omega_n)=
(1-\mathcal{G}^{(0)}_{\sigma}(i\omega_n)V^{\mathcal R}_{\alpha\sigma})^{-1}\mathcal{G}^{(0)}_{\sigma}(i\omega_n)$, with local BCS Green function,
\begin{align}
\mathcal{G}^{(0)}_{\sigma,ij}(i\omega_{n})=\pi\nu_F
\frac{-i\omega_n\tau^{0}_{ij}+\sigma\Delta\tau^{1}_{ij}}{\sqrt{|\Delta|^2-(i\omega_n)^2}}
\end{align}
and spin-dependent tunnelling self-energy
\begin{equation}
\Sigma_{\alpha\alpha',\sigma}=\begin{pmatrix}
0 && V^{\mathcal T}_{\sigma} \\ {V^{\mathcal T}_{\sigma}}^{\dagger} && 0
\end{pmatrix}_{\!\!\alpha\alpha'}.
\label{self_energy}
\end{equation}
The retarded/advanced Green functions are obtained by analytical continuation, and restricting our attention to linear response of a thermal equilibrium state, the lesser/greater (Nambu) correlation functions follow simply from the fluctuation-dissipation theorem: 
\begin{align}
G^{<}_{\alpha\alpha'\!,\sigma}(\omega)&=i A_{\alpha\alpha'\!,\sigma}(\omega)n_F(\omega),\\
G^{>}_{\alpha\alpha'\!,\sigma}(\omega)&=i A_{\alpha\alpha'\!,\sigma}(\omega)\big(n_F(\omega)-1\big),
\end{align}
where $n_{F}(\omega)$ denotes the Fermi function, and lead-indexed Nambu-matrix spectral functions are given by
\begin{align}
A_{\alpha\alpha'\!,\sigma}(\omega)=i(G^R_{\alpha\alpha'\!,\sigma}(\omega)-G^A_{\alpha\alpha'\!,\sigma}(\omega)).
\end{align}

\subsection{Bound state energies}
The denominator of the retarded Green function is given by
\begin{align}\label{eq:Dfct}
D_\sigma(\omega+i\eta)=&\,(1+\chi u)\Delta^2-(1+u)(\omega+i\eta)^2\\
&\hspace*{10mm}+2g\sigma\omega\sqrt{\Delta^2-(\omega+i\eta)^2},\nonumber
\end{align}
with $u=w^2-g^2$ and $\chi=1-\sin^2(2\theta)\sin^2(\varphi/2).$ The roots of $D_\sigma(\omega+i\eta)$ determine the bound state energies to be~\cite{Kirsanskas2015Dec}
\begin{align}
E_{\pm\sigma}=&\frac{s_\pm\sigma\Delta}{\sqrt{(1+u)^2+4g^2}}\Big(2g^2+(1+\chi u)(1+u)\label{YSRBSenergies}\\
&\hspace{15mm}\pm2g\sqrt{g^2+u(1-\chi)(1+\chi u)}\Big)^{1/2},\nonumber
\end{align}
where $s_-=-\text{sign}(1+\chi u)$ and $s_+=\text{sign}(u)$. This is valid for general potential, and exchange cotunnelling amplitudes, $w$ and $g$, but the Schrieffer-Wolff transformation will always lead to a negative $u$ (cf. Eq.~\eqref{eq:gw}). Note that the $\chi$ defined here is not related to the susceptibility discussed above, but merely the short-hand notation chosen in Ref.~\onlinecite{Kirsanskas2015Dec}.

\begin{figure}[t]
\centering
\includegraphics[width=\linewidth]{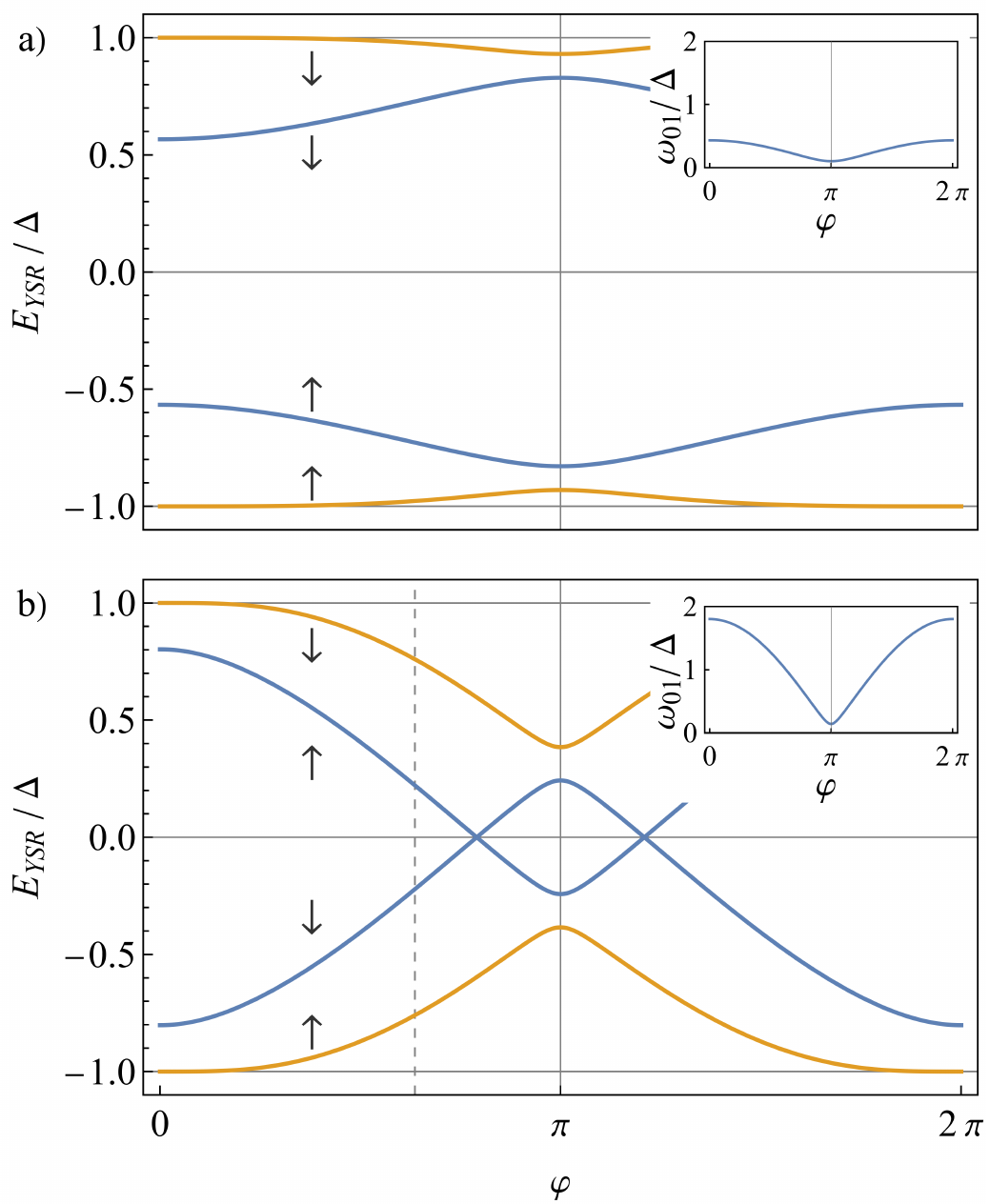}
\caption{YSR bound state energies plotted against the phase difference for $\Gamma_{R}/\Gamma_{L}=0.35/0.30$, corresponding to $\theta=\pi/3.81$, and gate valtage corresponding to $x=0.5$,  away from the particle-hole symmetric point. The energies $E_{+\sigma}$ ($E_{-\sigma}$) are shown in orange (blue). Sub-gap excitation spectrum (a) at weak coupling, $\Gamma/U=0.1$, and (b) at stronger coupling, $\Gamma/U=0.6$. The gray dashed line refers to Fig.~\ref{fig:admittance}, and the insets show the two-quasiparticle excitation energies,  $\omega_{01}$.}
\label{fig:Eysr}
\end{figure}

The spin-dependent cotunnelling term, with amplitude $\sigma JS$, breaks the spin degeneracy of the bound states, and we obtain four distinct bound state energies, $E_{\pm\sigma}$, 
which are plotted against phase difference in Fig.~\ref{fig:Eysr}. The $\pm$-index refers to the two channels involved in the problem. At zero phase difference only the minus states, $|-,\sigma\rangle$, have energies inside the gap, and these are the states originally found by Yu, Shiba, and Rusinov~\cite{Yu1965, Shiba1968, Rusinov1969}. The plus states, $|+,\sigma\rangle$, only enter the gap at a finite phase difference, and at $\varphi=\pi$, they become degenerate with the minus states of same spin, unless either particle-hole, or left-right coupling symmetry is broken. The quasiparticle spins indicated in the figure tacitly assume the dot spin to point up. The state $|-,\downarrow\rangle$ therefore corresponds to the dot spin being screened by $\hbar/2$, and would really have been the YSR singlet state in an unpolarized treatment of the model~\cite{Bauer2007,Kirsanskas2015Dec}. For the weak couplings chosen in Fig.~\ref{fig:Eysr}(a), this is an excited state. At stronger coupling (Fig.~\ref{fig:Eysr}(b)), this state crosses zero energy, signifying a change in ground state from unscreened to screened. The lowest-lying excitation from the screened ground state, is the opposite spin partner, $|-,\uparrow\rangle$, corresponding to the unscreened spin doublet within an unpolarized treatment.

\begin{figure}[t!]
\centering
\includegraphics[width=0.95\linewidth]{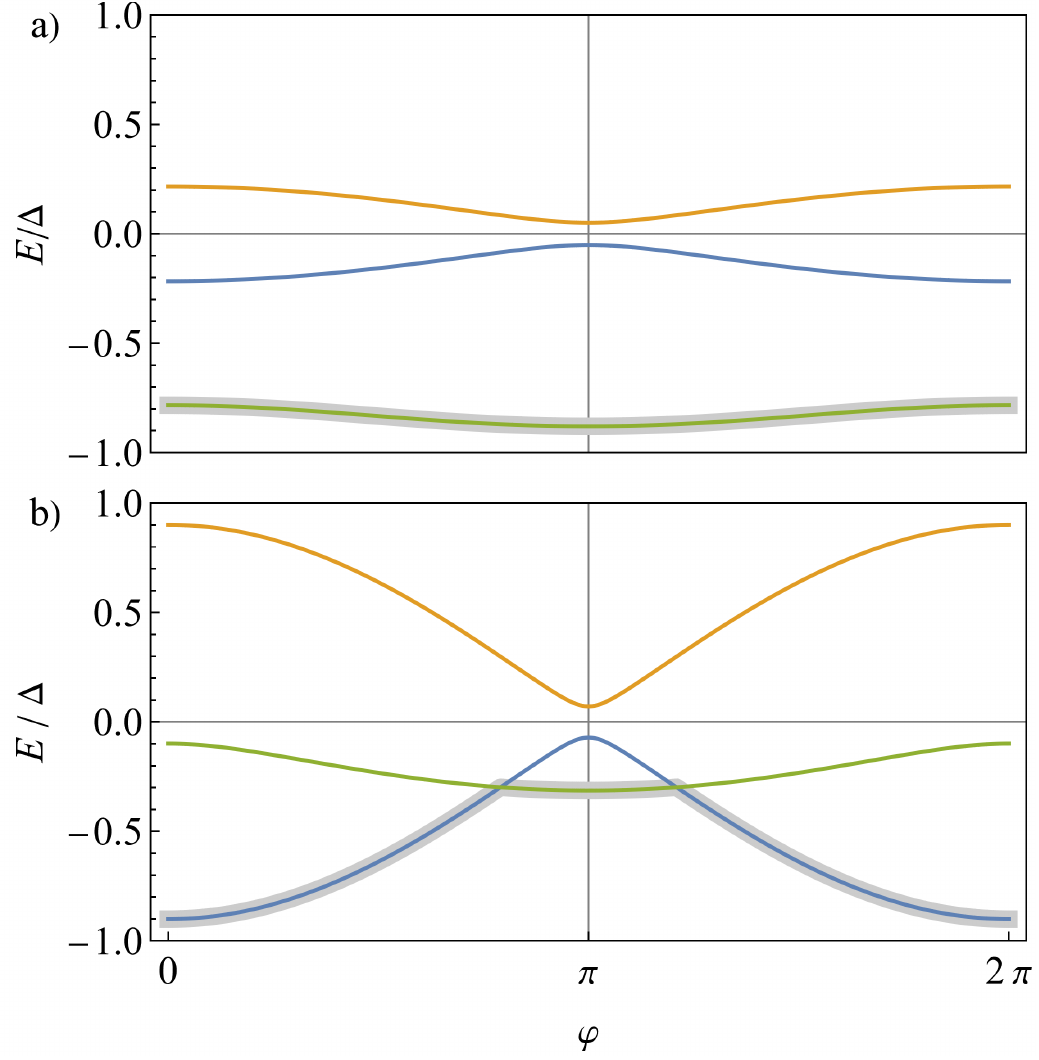}
\caption{Many-body eigenenergies, $E_{s,0}$ (blue), $E_{s,1}$ (orange) and $E_{u,0}$ (green), calculated within the polarized-spin approximation. The wide gray line indicates the ground state energy, $E_{GS}$.
Parameters in panels (a,b) are the same as in Fig.~\ref{fig:Eysr}(a,b)}
\label{fig:mbe4a}
\end{figure}

From the single-particle excitaion energies in Eq.~\eqref{YSRBSenergies}, we may infer the eigenenergies of the many-body eigenstates of this non-interacting polarized-spin cotunnelling Hamiltonian. For the two lowest-lying screened states, and the lowest-lying unscreened state, the eigenenergies can be found, up to a common constant off-set, as
\begin{align}\label{mben}
E_{s,0}=&(E_{+,\uparrow}+E_{-,\downarrow})/2,\nonumber\\
E_{s,1}=&(E_{-,\uparrow}+E_{+,\downarrow})/2=-E_{s,0},\\
E_{u,0}=&(E_{-,\uparrow}+E_{+,\uparrow})/2,\nonumber
\end{align}
utilizing that $E_{\mp,\sigma}=-E_{\mp,-\sigma}$. The two screened states are connected by spin-conserving two-particle excitation, as induced by the current operator, with energy $\omega_{01}=E_{s,1}-E_{s,0}=2E_{s,1}$. This excitation energy is shown as an inset in Figs.~\ref{fig:Eysr}(a,b). The screened, and unscreened states have different spin, and will only be connected by single-particle excitations with energies $E_{u,0}-E_{s,0}=E_{-,\uparrow}$, and $E_{u,0}-E_{s,1}=E_{+,\downarrow}$. In Fig.~\ref{fig:mbe4a}, we show the phase-dispersion of the eigenenergies~\eqref{mben}, for the same parameters as were used in Fig.~\ref{fig:Eysr}, with the corresponding ground state energy, 
\begin{align}\label{EGS}
E_{GS}=-\sum_{n=\pm,\sigma}|E_{n,\sigma}|/4,
\end{align}
indicated by the wide grey line.

\subsection{Equilibrium supercurrent}
The cotunnel current operator, $\hat{I}=\partial_{\phi}\hat{H}$, now takes the following form:
\begin{align}\label{currentop}
\hat{I}&=\frac{ie}{\hbar}\sum_{kk'\sigma}
\left[
\psi^{\dagger}_{Lk\sigma,i}\tau^{3}_{ij}
V^{\mathcal T}_{\sigma,ij}
\psi^{}_{Rk'\sigma,j}\right.\\
&\left.\hspace*{30mm}
-\psi^{\dagger}_{Rk\sigma,i}\tau^{3}_{ij}
{V_{\sigma,ij}^{\mathcal T}}^{\!\!\!\ast}
\psi^{}_{Lk'\sigma,j}\right],\nonumber
\end{align}
with ground-state expectation value given by 
\begin{align}
\langle\hat{I}\rangle=&-\frac{e}{2\hbar}\sum_{\sigma}\int_{-\infty}^\infty
\frac{d\omega}{2\pi}\nonumber\\
&\hspace*{5mm}\times\text{Tr}\big[\tau^3
\big(V_\sigma^{\mathcal T}G^{<}_{RL,\sigma}(\omega)
-{V_{\sigma}^{\mathcal T}}^{\dagger}G^{<}_{LR,\sigma}(\omega)\big)
\big]\nonumber\\
=&\frac{e}{\hbar}\Delta^2(g^2-w^2)\sin^2(2\theta)\sin(\varphi)\nonumber\\
&\hspace*{5mm}\times\sum_{\sigma}\int_{-\infty}^{0}
\frac{d\omega}{2\pi}\text{Im}\Bigg(\frac{1}{D_\sigma(\omega+i\eta)}\Bigg)\label{IfrominvD},
\end{align}
in terms of the retarded denominator~\eqref{eq:Dfct}. The integral can be split into a contribution from the bound states and a contribution from the continuum. 

Outside the gap ($\omega<-\Delta$), the imaginary part comes from the square root in Eq.~\eqref{eq:Dfct}. Since this term is odd in $\sigma$, the spin sum in Eq.~\eqref{IfrominvD} vanishes and the continuum therefore does not contribute to the current. As pointed out in Ref.~\cite{Kirsanskas2015Dec}, this is an artifact of the polarized-spin approximation, which misses a potentially important continuum contribution, which can only be captured by a full quantum mechanical treatment of the dot spin.

Inside the gap ($-\Delta<\omega<0$), the denominator can be expanded around its roots, whereby the $\eta\rightarrow 0_{+}$ limit yields the current carried by the occupied sub-gap states
\begin{align}
\langle\hat{I}\rangle=&\,-\frac{e}{2}\Delta^2(g^2-w^2)\sin^2(2\theta)\sin(\varphi)\!\sum_{n=\pm,\sigma}\frac{\theta(-E_{n\sigma})}{D'_{\sigma}(E_{n,\sigma})}\nonumber\\
=&\,2e\,\partial_{\varphi}E_{GS},
\end{align}
with ground state energy~\eqref{EGS}. The derivative $D_{\sigma}'(\omega)=\partial_{\omega}D_{\sigma}(\omega)$ has been evaluated at the eigenenergies, using the fact that 
\begin{align}
D_{\sigma}'(E_{n,\sigma})\partial_{\varphi}E_{n,\sigma} =&\,-\Delta^{2}u\,\partial_{\varphi}\chi\\
=&\,-\frac{1}{2}\Delta^{2}(g^{2}-w^{2})\sin^{2}(2\theta)\sin(\varphi),\nonumber
\end{align}
which follows from differentiating the equation $D_{\sigma}(E_{n,\sigma})=0$ with respect to $\varphi$. 

To leading order in $g$ and $w$, this supercurrent becomes
\begin{align}
\langle\hat{I}\rangle_{0}\approx-\frac{e\Delta}{2}(g^{2}-w^{2})\sin^{2}(2\theta)\sin(\varphi),
\end{align}
which, since $g^{2}>w^{2}$ for all gate-voltages, displays the expected $\boldsymbol{\pi}$ junction character of the spinful cotunnel junction at weak coupling~\cite{Kulik1966Apr}. Within the classical-spin approximation, the normal state tunnel resistance is given by $R_{N}=\pi\hbar/(e^{2}\tau)$ with transmission
\begin{equation}\label{tau}
\tau=\frac{\sin^2(2\theta)}{2}\Big(\frac{(w-g)^2}{1+(w-g)^2}+\frac{(w+g)^2}{1+(w+g)^2}\Big),
\end{equation}
and the weak-coupling supercurrent may be expressed as
\begin{align}
\langle\hat{I}\rangle\approx\frac{\pi\Delta}{2eR_{N}}\left(\frac{w^{2}-g^{2}}{w^{2}+g^{2}}\right)\sin(\varphi),
\end{align}
which takes the same form as Kulik's generalization~\cite{Kulik1966Apr} of the Ambegaokar-Baratoff result~\cite{Ambegaokar1963Jun} to including spin-flip tunnelling. Note that the polarized-spin approximation employed here involves merely spin-dependent, but not spin-flip cotunnelling. The transmission~\eqref{tau} therefore only really shares the weak, and the strong-coupling limits with that of the full unpolarized Kondo problem including spin flip terms. In the strong-coupling limit, $\Gamma\gg U$, also the Schrieffer-Wolff transformation looses its validity, but we note that in the limit of $g,w\rightarrow\infty$ with $g\neq w$, the sub-gap energies found within this spin-dependent cotunnelling model remain well-defined and become pairwise degenerate with
\begin{align}
-E_{+\uparrow}=-E_{-\downarrow}=E_{+\downarrow}=E_{-\uparrow}=\Delta\sqrt{\chi}.
\end{align}
This corresponds to the result for a weak link~\cite{Beenakker1991Dec, Furusaki1999May, Kos2013May} and the sub-gap states carry the supercurrent
\begin{align}\label{eq:strongsc}
\langle\hat{I}\rangle=\frac{e\Delta}{2\hbar}\frac{\sin^{2}(2\theta)\sin(\varphi)}{\sqrt{1-\sin^{2}(2\theta)\sin^{2}(\varphi/2)}},
\end{align}
corresponding to a $\boldsymbol{0}$ junction. This is exactly the zero-temperature result found in Ref.~\onlinecite{Glazman1989} for the supercurrent through a Kondo impurity in a tunnel barrier, with Kondo temperature, $T_{K}$, much larger than the gap, $\Delta$. In this limit, the fully screened Kondo impurity (here Coulomb-blockaded spinful quantum dot) behaves as a resonant level of width $T_{K}\gg\Delta$, for which the supercurrent is indeed carried entirely by its sub-gap states~\cite{Beenakker1992, Martin-Rodero2011Dec}, and the non-sinusoidal current-phase relation attains its maximum (critical) value,
\begin{align}
I_{C}=\frac{e\Delta}{\hbar}\left(1-\cos(2\theta)\right),
\end{align}
at $\varphi=\arccos\left[1-2(1-\cos(2\theta))/\sin^{2}(2\theta)\right]$. 

At intermediate coupling strengths, the cotunnel junction exhibits also a $\boldsymbol{\pi^\prime}$, and a $\boldsymbol{0^\prime}$ phase in which the current-phase relation is discontinuous and the inductive response therefore very sensitive to phase changes. At $\Gamma\ll U$, the cotunnel junction constitutes a $\boldsymbol{\pi}$ junction, and increasing $\Gamma/U$, it crosses through a $\boldsymbol{\pi^\prime}$, into a $\boldsymbol{0^\prime}$ phase and finally ends up as a $\boldsymbol{0}$ junction for $\Gamma\gg U$, unless the system is fine-tuned to $\theta=\pi/4$ and $\epsilon_{d}=-U/2$. This general trend is captured already within the polarized-spin approximation~\cite{Kirsanskas2015Dec}, but the detailed phase boundaries are different from those obtained in an unpolarized treatment including the spin-flip terms~\cite{Rozhkov1999Mar, Tanaka2007May, Zonda2015Mar}.

\begin{figure*}[t!]
\centering
\includegraphics[width=\linewidth]{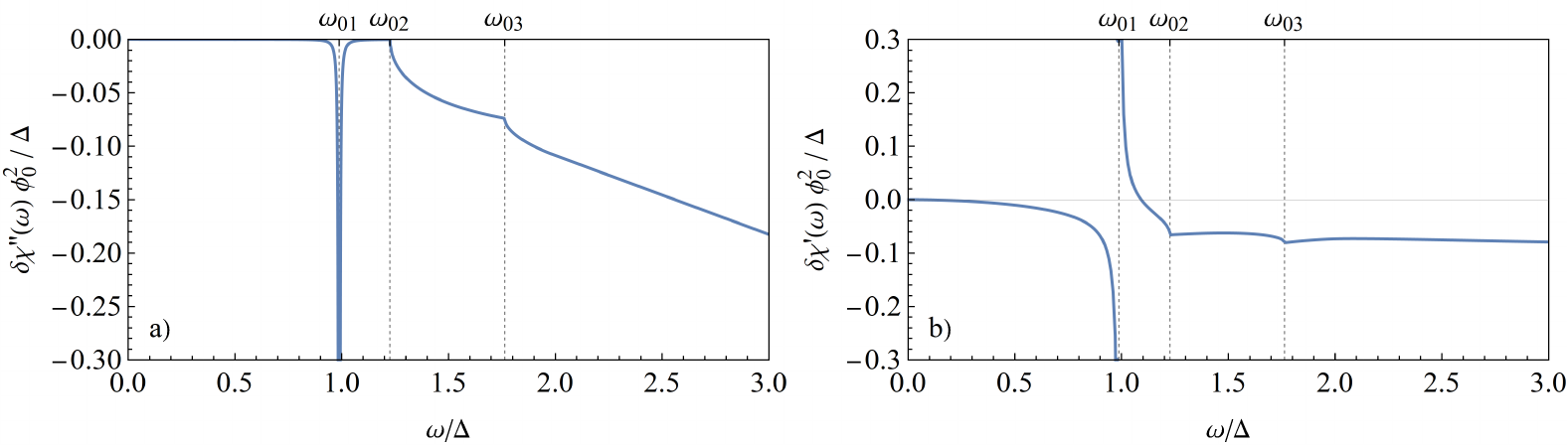}
\caption{(a) Imaginary, and (b) real part of $\delta\chi(\omega)$ as a function of frequency, calculated within the polarized-spin approximation. The dip at frequency $\omega_{01}=E_{+\downarrow}+E_{-\uparrow}$, reflects the creation of two quasiparticles of opposite spin in the sub-gap states. Creating instead a single quasiparticle in the continuum leads to the two threshold frequencies $\omega_{02}=\Delta+E_{-\uparrow}$ and $\omega_{03}=\Delta+E_{+\downarrow}$. Parameters are the same as used in Fig.~\ref{fig:Eysr}(b) ($x=0.5,$ $\theta=\pi/3.81,$ $\Gamma=0.6U$), and with $\varphi=2$ (as indicated by the gray dashed line in Fig.~\ref{fig:Eysr}(b)) corresponding to a screened ground state. Note that the $\delta$-function dip in $\delta\chi^{\prime\prime}(\omega)$ at $\omega=\omega_{01}$ is plotted in (a) as a Lorentzian of width $\eta=10^{-4}\Delta$. This sharp dip, would not show up in a corresponding plot of $\delta\chi^{\prime\prime}(\omega)$ in the unscreened ground state at $\varphi=\pi$.}
\label{fig:admittance}
\end{figure*}
\subsection{Linear response}\label{sec:LinRes}
Henceforth we employ units in which $\hbar=1$. The linear response of the cotunnel junction supporting YSR states is calculated as described in Sec.~\ref{sec:lrac}, using Eq.~\eqref{eq:respf}, without the subscript \textit{n}, since we shall only consider the linear response of the system in its ground state. Within the polarized-spin cotunnelling model with Hamiltonian ~\eqref{Heff}, the response function reduces to a simple convolution of two single-particle Green functions
\begin{align}
\delta\chi''(\omega)&=-\frac{e^2}{4\pi}\sum_\sigma \int_0^\omega d\epsilon\label{YSR_ReY}\\
&\hspace*{7mm}\times\text{Tr} \big[ \tau^3{V_\sigma^{\mathcal T}}^{\dagger} A_{LL,\sigma}(\epsilon-\omega)
\tau^3V_\sigma^{\mathcal T} A_{RR,\sigma}(\epsilon)\nonumber\\
&\hspace*{14mm}-\tau^3V_\sigma^{\mathcal T} A_{RL,\sigma}(\epsilon-\omega)
\tau^3V_\sigma^{\mathcal T} A_{RL,\sigma}(\epsilon)\big].\nonumber
\end{align}
Since the response function, $\chi^R(\omega)$, is analytic in the upper-half plane and $\chi^R(0)$ is real, $\delta\chi'(\omega)$ can be obtained from the Cauchy integral formula as 
\begin{align}
\delta\chi'(\omega)&=\frac{1}{\pi}\int_{-\infty}^\infty d\varepsilon\,\delta\chi''(\varepsilon)\left(\frac{1}{\varepsilon-\omega}-\frac{1}{\varepsilon}\right)\nonumber\\
&=\frac{1}{\pi}\int_{0}^\infty d\varepsilon\,\delta\chi''(\varepsilon)\left(\frac{1}{\varepsilon-\omega}+\frac{1}{\varepsilon+\omega}-\frac{2}{\varepsilon}\right)
\label{KramersKronig},
\end{align}
which is reminiscent of the usual Kramers-Kronig relations. In this way, the real part of $\chi_{II}(\omega)$ can finally be obtained as
\begin{equation}
\chi_{II}'(\omega)=\partial^2_\phi E_{\text{GS}}+\delta\chi'(\omega).  
\label{RealPartChi}
\end{equation}

\subsection{Results}\label{sec:results}


We perform the integration in Eq.~\eqref{YSR_ReY} numerically for different values of coupling strength, asymmetry and gate voltage. An example is seen in Fig.~\ref{fig:admittance} for a case of intermediate coupling strength and an arbitrary value of the phase for which the system is in the screened ground state, i.e. to the left of the crossing of the energies $E_{-\sigma}$ in Fig.~\ref{fig:Eysr}(b). The plot of $\delta\chi''(\omega)$ shows three distinct threshold frequencies, each corresponding to a specific absorption process. A $\delta$-function dip is observed at the frequency $\omega_{01}=E_{+\downarrow}+E_{-\uparrow},$ at which two quasiparticles of opposite spin are created in the sub-gap states. This corresponds to a spin-conserving transition from the ground state with a screening quasiparticle in the minus channel to an excited state with a screening quasiparticle in the plus channel. In general, the current operator only connects states which differ by two quasiparticles of opposite spin. Therefore, the resonant contribution is absent in the unscreened ground state, where conservation of spin requires the final state to lie in the continuum. The two absorption processes with threshold frequencies $\omega_{02}=\Delta+E_{-\uparrow}$ and $\omega_{03}=\Delta+E_{+\downarrow},$ lead to the creation of a quasiparticle in one of the sub-gap states and another quasiparticle in the superconducting continuum.
 
From Eq.~\eqref{YSR_ReY} an analytical expression for the resonant contribution to $\delta\chi''(\omega)=\chi''(\omega)$ can be obtained (cf. Appendix~\ref{Appf}) by considering the part of the integration interval where $\varepsilon<\Delta$ and $\omega-\epsilon<\Delta.$ Without loss of generality we restrict our attention to positive frequencies $\omega>0,$ whereby
\begin{align}
\delta\chi_{\text{res}}''(\omega)&=-\pi|I_{01}|^{2}\delta(\omega-\omega_{01})\theta(-E_{-\downarrow}),
\label{eq:ResContImChi}
\end{align}
with  resonance frequency $\omega_{01}=E_{+\downarrow}+E_{-\uparrow}$ and norm-squared current matrix element
\begin{align}\label{eq:currmatelem}
|I_{01}|^{2}&=\!\!\sum_{\sigma; i,j=\pm}\!\!\!\frac{f_\sigma(E_{i\sigma},E_{i\sigma}-E_{j\sigma})}{D'(E_{i\sigma})D'(E_{j\sigma})}\theta(E_{i\sigma})\theta(-E_{j\sigma}).
\end{align}
The full expression for $f_\sigma(\epsilon,\omega)$ is rather lengthy and is provided in appendix~\ref{Appf}.
\begin{figure*}[t]
\centering
\includegraphics[width=0.48\linewidth]{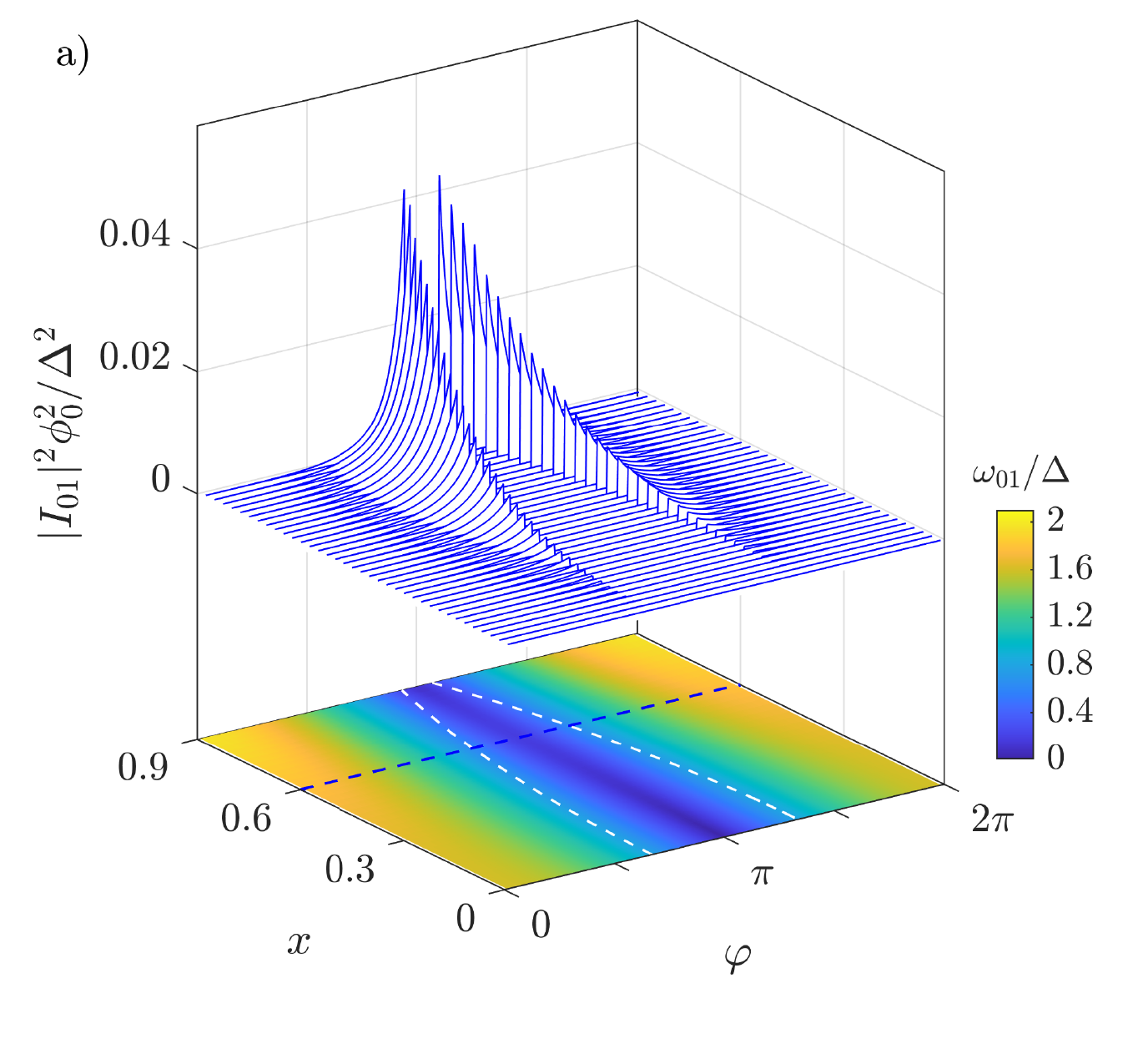}\quad
\includegraphics[width=0.48\linewidth]{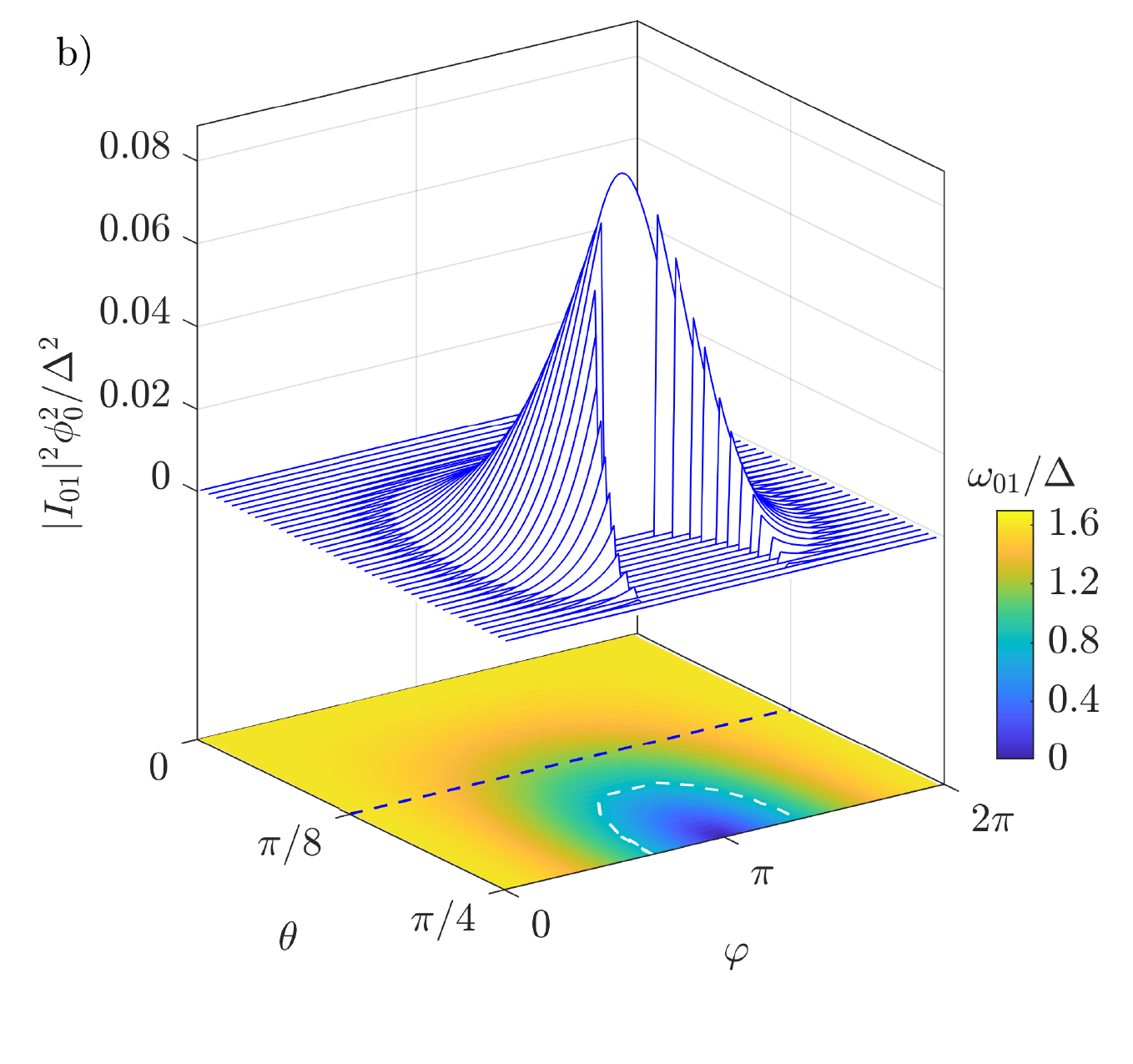}\\
\caption{Current matrix element determining $\chi_{II}''$ at resonance (blue curves) together with the corresponding transition energy (density plot) plotted as a function of phase bias and, respectively, (a) particle-hole asymmetry, $x$, at  symmetric coupling $\theta=\pi/4$ or (b) coupling asymmetry, $\theta$, at the particle-hole symmetric point ($x=0$). In both plots the coupling strength is $\Gamma/U=0.5$. White dashed lines delimit the region with the unscreened ground state for which the matrix element for microwave absorption is zero. Inside this unscreened region, the excitation energy refers to a transition between the excited screened states. Blue dashed lines indicate the cuts at which we show $\chi_{II}'$ in Figs.~\ref{fig:rechipolspinx06} and~\ref{fig:rechipolspinx0}.}
\label{fig:YSRY3}
\end{figure*}

From Eq.~\eqref{eq:ResContImChi} we see immediately, that the resonant contribution vanishes in the unscreened ground state, where the energy $E_{-\downarrow}$ is positive. The matrix element of the current operator between the plus, and minus screened states gives the magnitude of $\delta\chi''_{\text{res}}$ and is plotted in Fig.~\ref{fig:YSRY3} along with the resonance frequency as a function of phase difference and, respectively, junction coupling asymmetry and particle-hole asymmetry. From these plots, we also learn that the absorption is largest when the bound states are deep inside the gap, resulting in a resonance at low energy. 

In the limit where $g\to 0$, our cotunnelling model corresponds to a weak link~\cite{Kos2013May}, and from Eq.~\eqref{YSRBSenergies} we recover the expected bound state energies, $\pm E_{A}$, with $E_{A}=\Delta\sqrt{1-\tau\sin^{2}(\varphi/2)}$ and transmission given by Eq.~\eqref{tau} with $g=0$. In this limit, one finds that
\begin{align}
f_{\sigma}(E_{A},2E_{A})&=2(1+w^{2})^{2}(\Delta^{2}-E_{A}^{2})\nonumber\\
&\hspace*{20mm}\times(E_{A}^{2}-\Delta^{2}\cos^{2}(\varphi/2)),
\end{align}
and since $D'_{\sigma}(\omega)=-2\omega(1+w^{2})$, the real part of the ($\omega>0$) admittance
becomes
\begin{align}\label{eq:Ywl}
Y_{\text{res}}'(\omega)&=-\frac{1}{\omega}\delta\chi_{\text{res}}''(\omega>0)\nonumber\\
&=\frac{\pi^{2}G_{N}(\Delta^{2}-E_{A}^{2})(E_{A}^{2}-\Delta^{2}\cos^{2}(\varphi/2))}{2E_{A}^{3}}\nonumber\\ &\hspace*{40mm}\times\delta(\omega-2E_A),
\end{align}
corresponding to the resonant contribution for a weak link, $Y^{(0)}_{3}(\omega)$, found by Kos et al.~\cite{Kos2013May}.

Likewise, with $w=0$, the strong coupling limit, $g\to\infty$ of our polarized-spin exchange-cotunnelling model again corresponds to a weak link with bound states at $\pm E_{A}$ with transmission $\tau=\sin^{2}(2\theta)$. Expanding Eq.~\eqref{eq:currmatelem}, 
one again confirms the weak-link result in Eq.~\eqref{eq:Ywl}. This limit obviously goes beyond the validity of the Schrieffer-Wolff transformation used to establish the exchange-cotunnelling model in the first place, but just like this limit was found to capture the correct supercurrent (Eq.~\eqref{eq:strongsc}) for an odd-occupied quantum dot at the particle-hole symmetric point with $T_{K}\gg\Delta$, the same may hold true for the resonant admittance. At this point, however, this remains only an asymptotic statement about the cotunnelling model itself.



The real part of $\chi_{II}(\omega)$ which yields the dispersive shift of the resonance frequency can be calculated using Eqs.~\eqref{KramersKronig} and~\eqref{eq:ResContImChi}. The integral can be split into three parts by partitioning the imaginary part as follows
\begin{align}\label{eq:imchi}
\delta\chi''(\omega)=\left\{
\begin{array}{ll}
\delta\chi''_{\text{res}}(\omega) &,\,\,0<\omega<\Delta+|E_{-\uparrow}|, \\ \\
\delta\chi''_{\text{cont}}(\omega) &,\,\,\Delta+|E_{-\uparrow}|<\omega<\omega_{c}, \\ \\
\delta\chi''_{\infty}(\omega) &,\,\,\omega_{c}<\omega<\infty,
\end{array}
\right.
\end{align}
where $\omega_{c}$ is a cutoff frequency chosen of the order of $2\Delta$, beyond which $\delta\chi''(\omega)$ decreases linearly with frequency as $\delta\chi''(\omega>\omega_{c})\approx-e^{2}\omega\tau/\pi\equiv \delta\chi''_{\infty}(\omega)$, as long as frequencies are much smaller than the lead electron bandwidth, which is tacitly assumed throughout.

In this way, the real part, $\delta\chi'(\omega)$, is written as the sum of three terms. The first term arises from the resonant contribution in Eq.~\eqref{eq:ResContImChi} and takes the following form
\begin{align}\label{chires}
\delta\chi'_{\mathrm{res}}(\omega)&=\frac{2|I_{01}|^{2}}{\omega_{01}}\frac{\omega^{2}}{\omega^{2}-\omega_{01}^{2}}\theta(-E_{-\downarrow}).
\end{align}
The second term, $\delta\chi_{\text{cont}}'(\omega)$, is the contribution involving the continuum of states with energies above the gap, and it is calculated numerically using Eq.~\eqref{YSR_ReY} up to the cutoff frequency, $\omega_c$. Finally, the error introduced by the finite cutoff frequency is easily calculated as
\begin{equation}\label{chiinf}
\delta\chi_\infty'(\omega)=\frac{e^2\tau}{\pi^2}\omega\log\left| \frac{\omega_c-\omega}{\omega_c+\omega}\right|.
\end{equation}
Adding up these three terms together with the inductive contribution, Eq.~\eqref{RealPartChi} takes the following form:
\begin{align}
\chi_{II}'(\omega)&=\partial^{2}_{\phi} E_{\text{GS}}+\delta\chi_{\text{res}}'(\omega)+\delta\chi_{\text{cont}}'(\omega)+\delta\chi_{\infty}'(\omega).
\end{align}
\begin{figure}[ht]
\centering
\includegraphics[width=\linewidth]{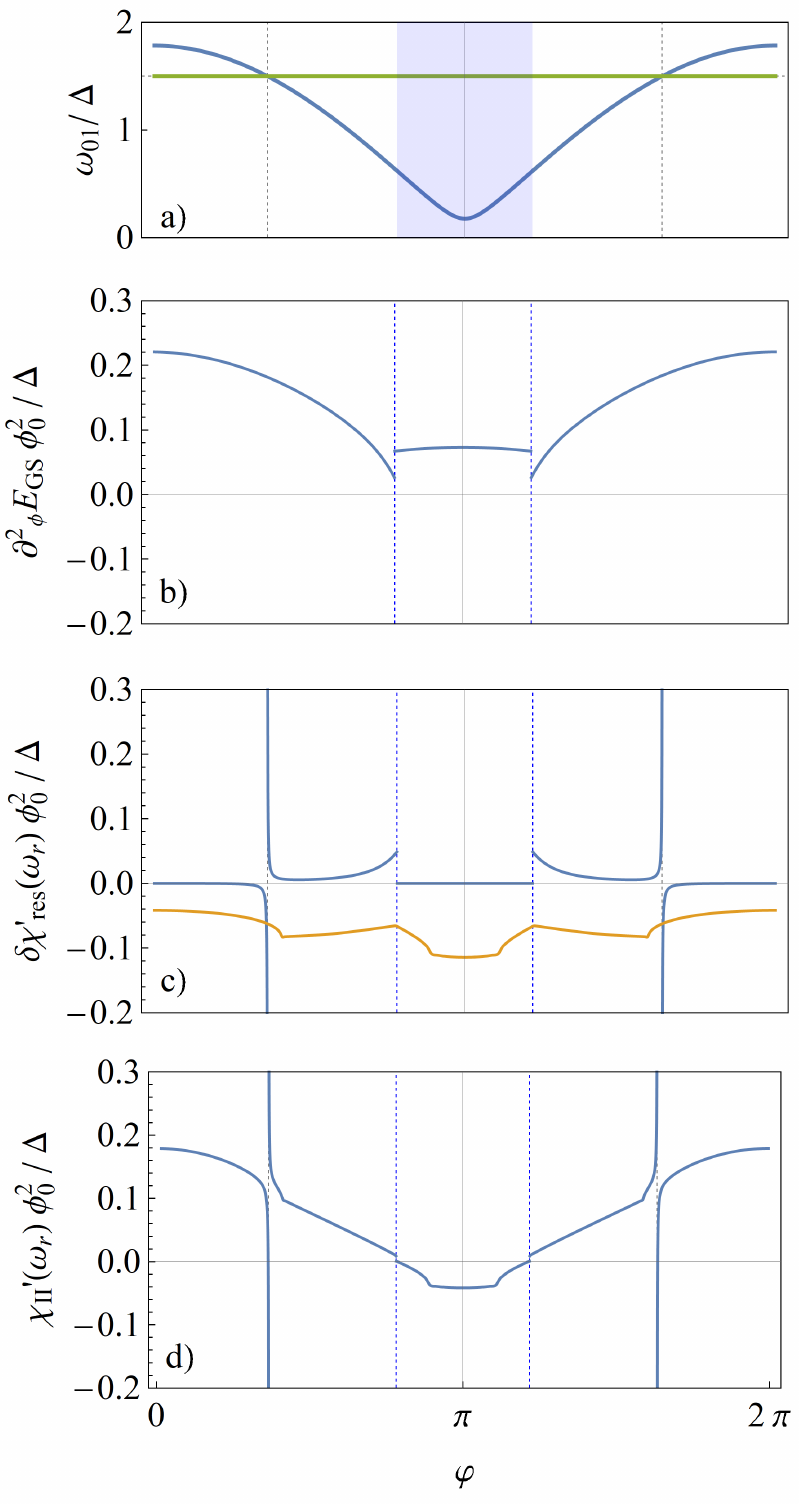}
\caption{(a) Two-quasiparticle excitation energy corresponding to a transition between the two screened states. The shaded region and blue vertical dashed lines delimit the region around $\varphi=\pi$ in which the ground state is unscreened. (b) Inductive, (c) resonant (blue line) and continuum (orange line) contributions to the total real part of the response function (d). Parameters are $\Gamma/U=0.5, x=0.6, \theta=\pi/4, \omega_{r}=1.5\Delta $, corresponding to the blue dashed line in Fig.~\ref{fig:YSRY3}(a)}
\label{fig:rechipolspinx06}
\end{figure}
\begin{figure}[ht]
\centering
\includegraphics[width=\linewidth]{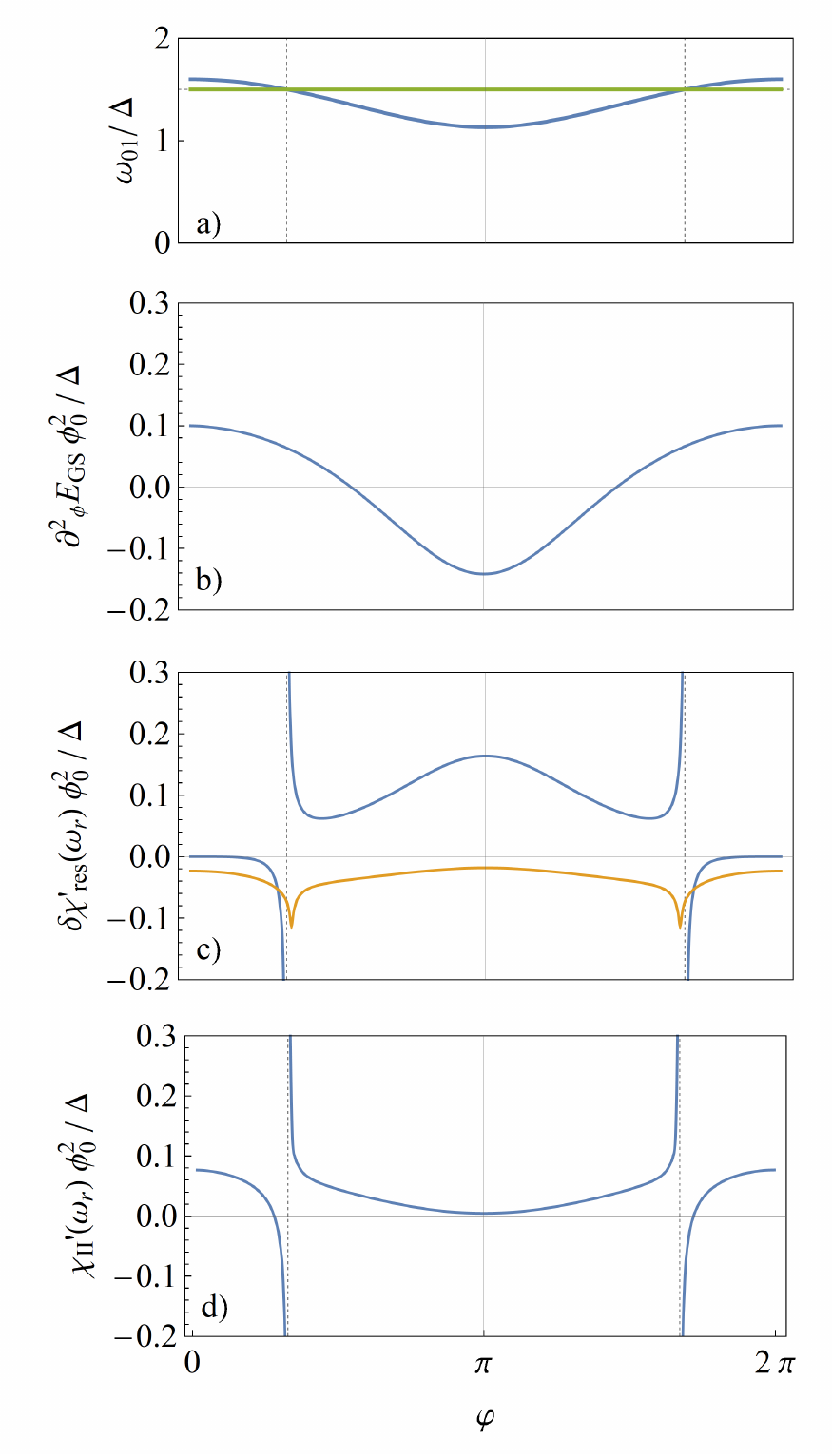}
\caption{(a) Two-quasiparticle excitation energy corresponding to a transition between the two screened states. (b) Inductive, (c) resonant (blue line) and continuum (orange line) contributions to the total real part of the response function (d). Parameters are $\Gamma/U=0.5, x=0, \theta=\pi/8, \omega_{r}=1.5\Delta$, corresponding to the blue dashed line in Fig.~\ref{fig:YSRY3}(b).}
\label{fig:rechipolspinx0}
\end{figure}


In Figs.~\ref{fig:rechipolspinx06} and~\ref{fig:rechipolspinx0} we plot $\chi_{II}'(\omega)$ and its individual terms for a fixed frequency as a function of the phase difference for parameters corresponding to the cuts in Fig.~\ref{fig:YSRY3}, indicated by blue dashed lines, and for one extra set of parameters. Figs.~\Cref{fig:rechipolspinx06,fig:rechipolspinx0}(a) show the two-quasiparticle excitation energy corresponding to a transition between the two screened states, where the region with unscreened ground state is marked in blue. 

Figs.~\ref{fig:rechipolspinx06} and~\ref{fig:rechipolspinx0} illustrate a general feature of linear microwave response of YSR states, that regions with a two-quasiparticle excitation energy far below $\Delta$ will have an unscreened ground state, and therefore no resonant excitation directly from the ground state. The screened ground state typically only offers a high excitation energy. With a typical Al gap of $\Delta\approx$ 0.2 meV, corresponding to 48 GHz, a resonator frequency close to $\Delta$ is practically unattainable. As we shall discuss in Sec.~\ref{sec:Noneq}, this problem can be overcome when the system is prepared in an excited state.  


Figs.~\cref{fig:rechipolspinx06,fig:rechipolspinx0}(b) show the inductive response, which constitutes a substantial contribution to the non-resonant response. At the transition from the screened, to the unscreened ground state (in Fig.~\ref{fig:rechipolspinx06}), the equilibrium supercurrent exhibits a sharp jump~\cite{Kirsanskas2015Dec,Rozhkov1999Mar}, which strictly speaking will give rise to a diverging phase-derivative, i.e. zero inductance, which has been cut out manually in these pots. These potentially large sharp dips in $\chi_{II}'(\omega)$ at the transition will be smeared by mixing of the two competing ground states, either by temperature or by quasiparticle poisoning, and their manifestation in experiment will rely entirely on the non-equilibrium dynamics within the given device. If, however, the measurement is carried out faster than the system switches between these competing ground states, the system will stay on a given supercurrent branch, now in an excited state, and exhibit a smooth inductive contribution.

In Figs.~\cref{fig:rechipolspinx06,fig:rechipolspinx0}(c), we show the resonant contribution together with the continuum contribution. Apart from the avoided crossing at resonance, one observes (in Fig.~\ref{fig:rechipolspinx06}) a pronounced cusp at the threshold between screened, and unscreened ground states. Also this feature will of course be smeared by mixing of the competing ground states or, in the case of a long enough ground state switching time, remain continuous across $\varphi=\pi$, much like in Fig.~\ref{fig:rechipolspinx0}(c) for the screened, and vanishing altogether for the unscreened state. 

Finally, Figs.~\cref{fig:rechipolspinx06,fig:rechipolspinx0}(d) show the total real part of the response function, which determines the frequency shift of the resonator according to Eq.~\eqref{eq:shift}, and which takes the following form:
\begin{align}
\frac{\delta\omega_{r}}{\omega_{r}}&=\lambda^{2}\frac{\Delta}{\hbar\omega_{r}}\left(\chi'_{II}(\omega_{r})\phi_{0}^{2}/\Delta\right).
\end{align}
When the resonator is out of resonance with the excitation energy, $\omega_{01}$, all other factors in this expression combine to a number roughly of the order of 0.1, corresponding to dispersive shifts of the order of 10\% of the dimensionless inductive coupling constant, $\lambda^{2}$. In Refs.~\onlinecite{Metzger2021Jan, Hays2021Jul}, this was estimated to be $\lambda^{2}\sim 10^{-5}-10^{-3}$, whereas relative frequency shifts could be resolved only down to $10^{-3}$. Observing the details of the non-resonant shifts induced by YSR states should therefore be feasible with state of the art techniques.

Similarly, the dissipative (imaginary) part of the response function, corresponds to the energy absorption rate of the cotunnel junction and gives rise to the following damping rate of the resonator
\begin{align}
\frac{\gamma_{r}}{\omega_{r}}&=\lambda^{2}\frac{\Delta}{\hbar\omega_{r}}\left(-\delta\chi''(\omega_{r})\phi_{0}^{2}/\Delta\right).
\end{align}
As evident from Fig.~\ref{fig:admittance}, this comprises the resonant contribution at $\omega_{r}=\omega_{01}$ as well as the continuum contribution at $\omega_{r}>\omega_{02}$. As observed in Fig.~\ref{fig:YSRY3}, the ground-state matrix element, $|I_{01}|^{2}$, which governs the resonant contribution is only significant in a confined region around the transition between screened, and unscreened ground states. This is markedly different from the infinite-gap limit in Fig.~\ref{fig:infgaptheta}(b), where a smooth background proportional to $\cos^{2}(2\theta)$ reaches its maximum at $\theta=0$, where the quantum dot is coupled only to the left superconducting lead. This background term is also present with finite gap and small charging energy, $U\ll\Delta$, with a prefactor of $\Delta^{2}/(\Delta+\Gamma)^{2}$~\cite{Kurilovich2021May}, but is seen here to vanish for $U\gg\Delta,\Gamma$ where the strong charging energy suppresses all real dot-lead charge currents and leaves only an effective inter-lead cotunnelling current, which vanishes when decoupling one of the leads.

\section{Zero-bandwidth approximation}~\label{sec:ZBW}

Up to this point, we have treated the YSR states within the polarized-spin approximation, which may be directly relevant for larger adatom spins subject to magnetic anisotropy on a superconducting surface~\cite{vonOppen2021May, Zitko2011Feb}, but which merely amounts to a tractable approximation for describing quantum dot systems with $\Gamma\ll U$ and gated to be close to the particle-hole symmetric point ($|x|\ll 1$). In this subsection we employ instead the zero-bandwidth (ZBW) approximation~\cite{Allub1981Feb,Martin-Rodero2011Dec}, which is known to describe YSR states very well, up to a rescaling of $\Gamma$~\cite{Grove-Rasmussen2018Jun, EstradaSaldana2018Dec}. The ZBW approximation respects the spin-rotational invariance, whereby the states described above as screened states now become bona fide singlets, and the unscreened states become spin doublets. Furthermore, this approximation includes all charge, and spin fluctuations of the quantum dot, and may therefore be used to interpolate between the Schrieffer-Wolff transformed cotunnelling model ($\Gamma,\Delta\ll U$), with YSR sub-gap states induced by the magnetic moment on the dot, and the infinite-gap model ($\Gamma, U\ll\Delta$), where the dot is simply proximitized and the notion of YSR states becomes irrelevant. As we shall demonstrate, at least at a qualitative level, the salient features of the resonant microwave response presented above in both limits are captured rather well within the ZBW approximation.

\begin{figure*}[t]
\centering
\includegraphics[width=0.45\linewidth]{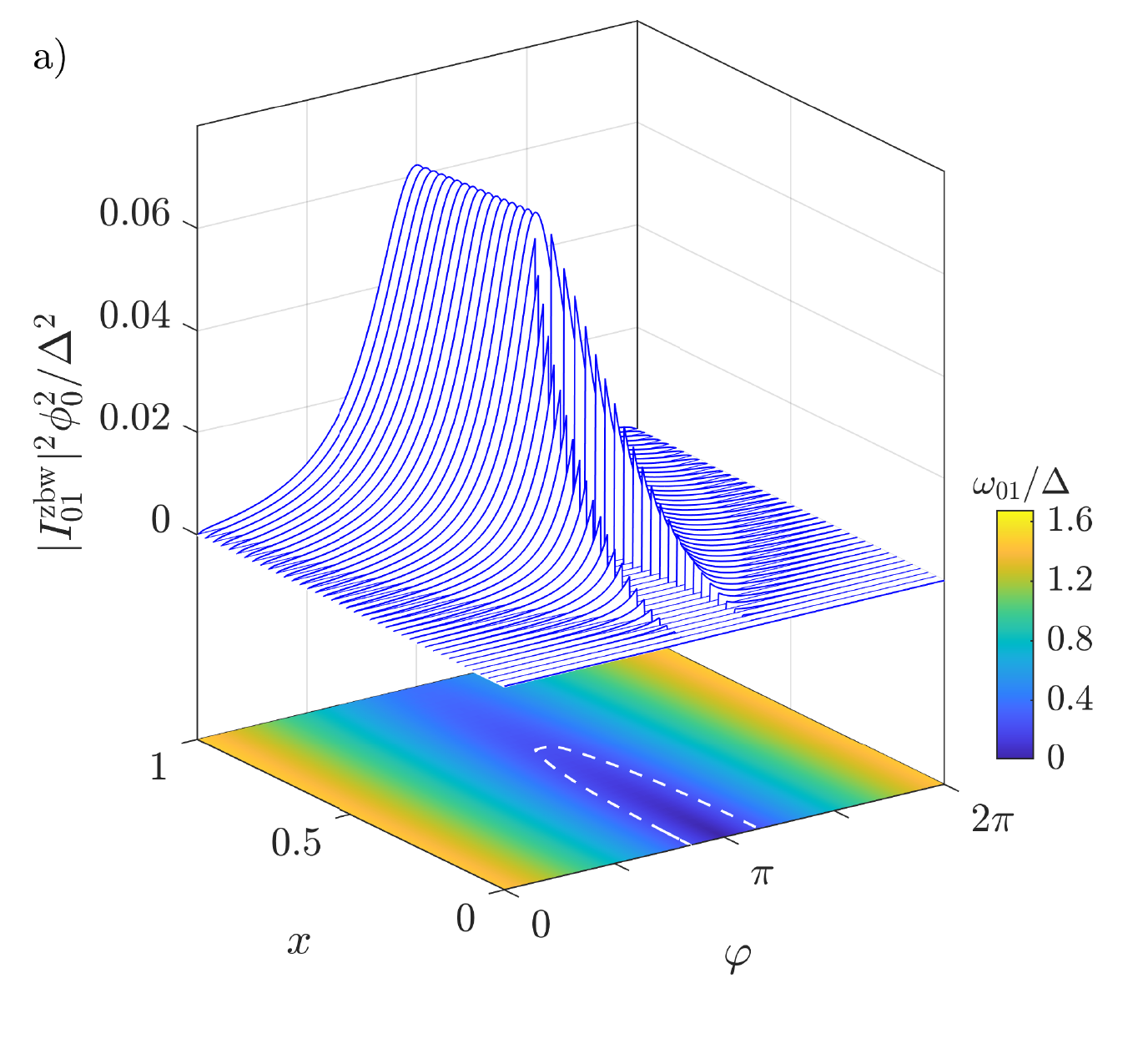}\quad
\includegraphics[width=0.45\linewidth]{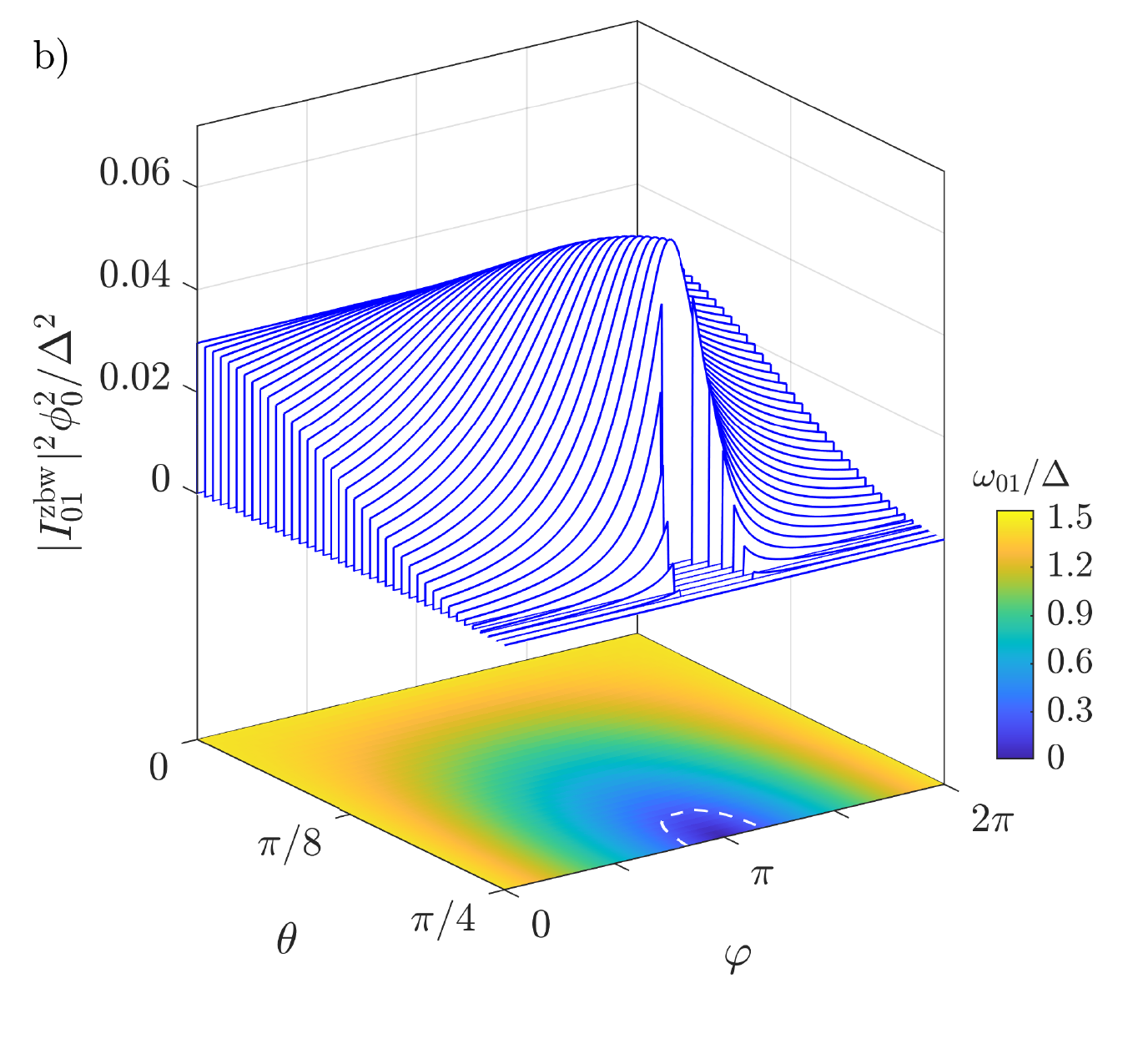}\\
\includegraphics[width=0.45\linewidth]{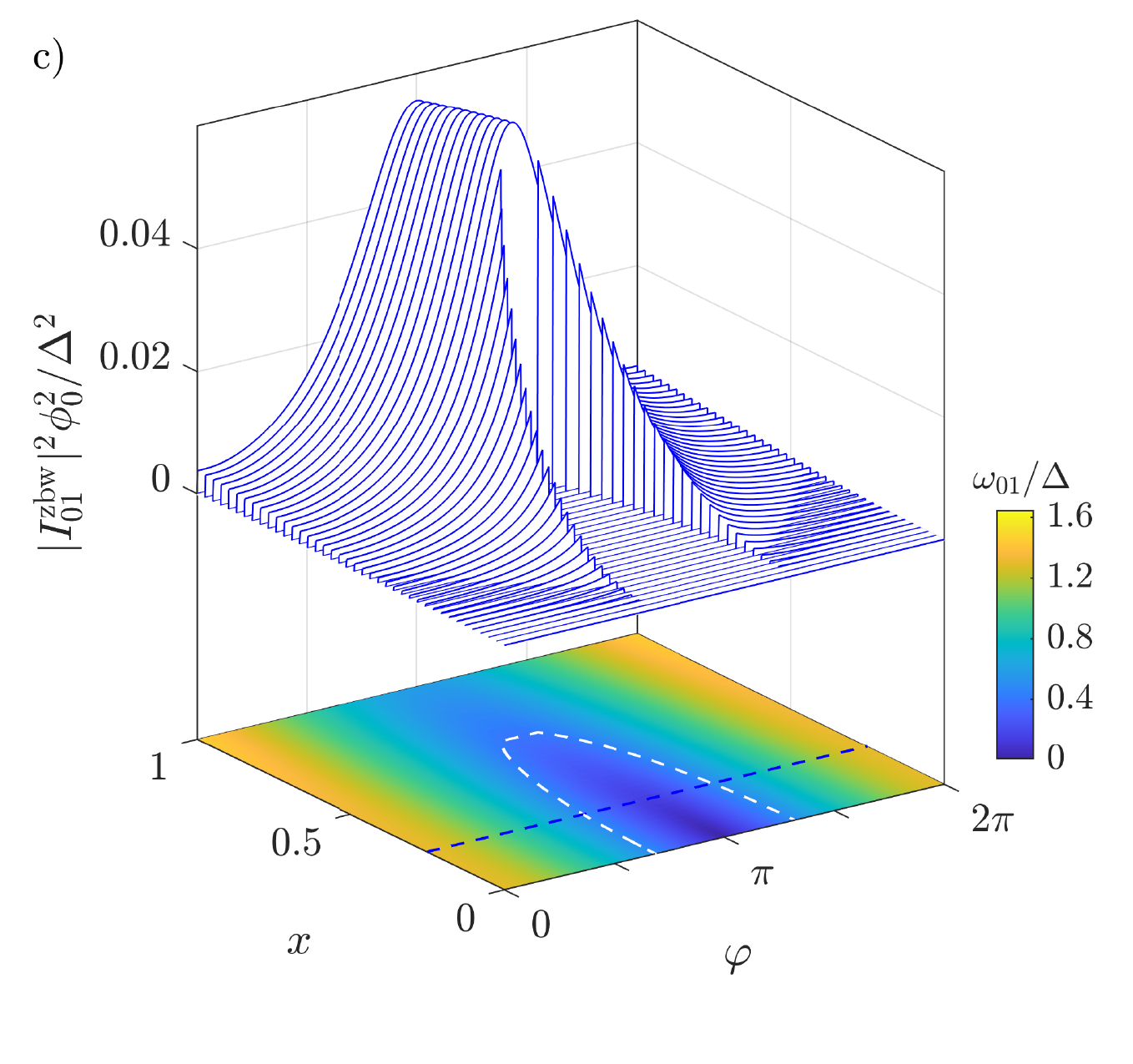}\quad
\includegraphics[width=0.45\linewidth]{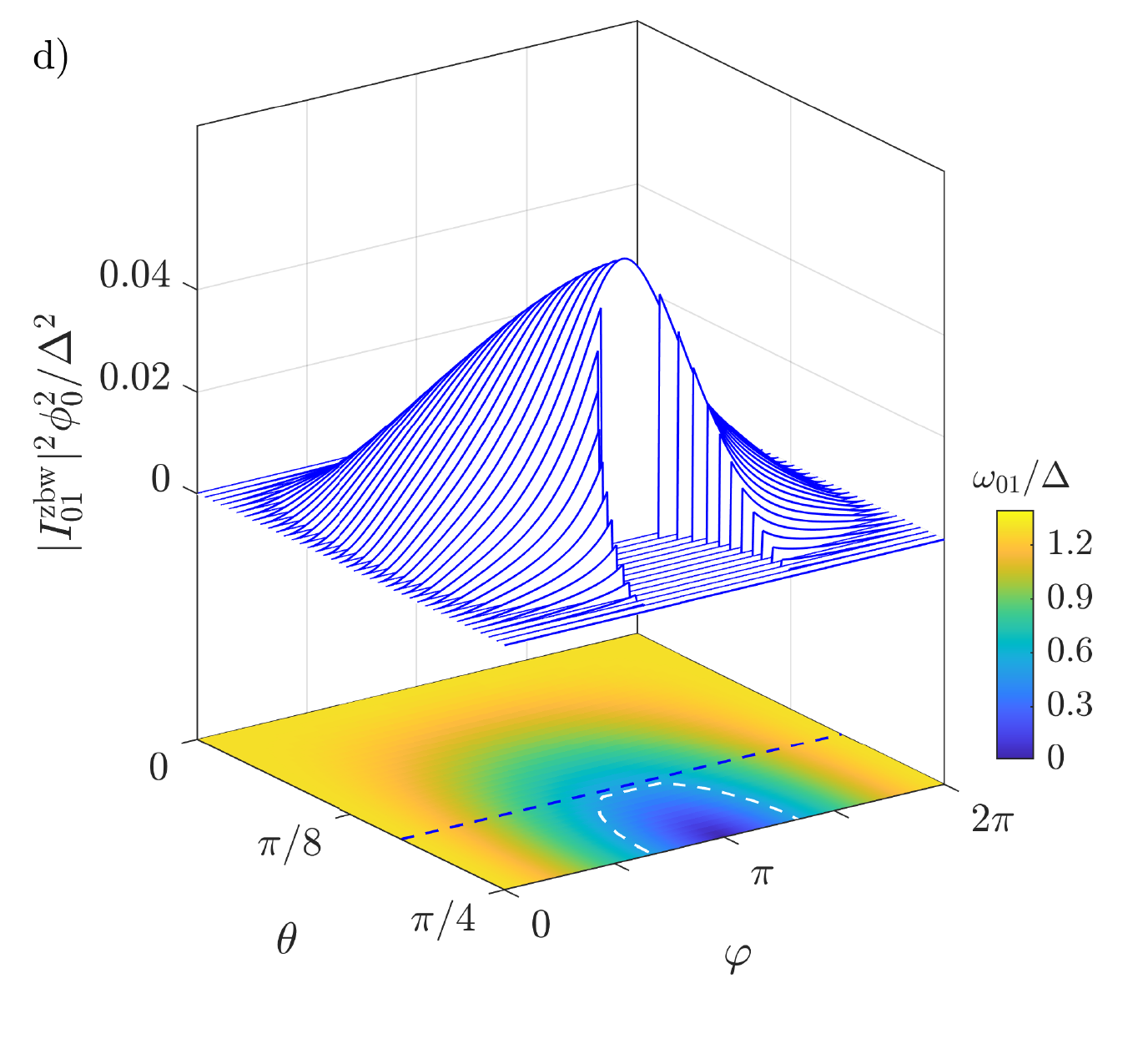}
\caption{Current matrix element determining $\chi_{II}''$ at resonance calculated within the ZBW model (blue curves) together with the corresponding transition energy (bottom) plotted as a function of phase bias and, respectively, particle-hole asymmetry, $x$, at symmetric coupling $\theta=\pi/4$, i.e. $\Gamma_{L}=\Gamma_{R}$ in (a) and (c) or coupling asymmetry, $\theta$, at particle-hole symmetry $x=0$, i.e. $\epsilon_{d}=-U/2$ in (b) and (d). In (a) and (b) $U=\Delta,$ in (c) and (d) $U=2\Delta,$ and $t=\Delta$ everywhere. White dashed lines delineate the region with doublet ground state, in which the dot spin is unscreened. Inside this doublet region the excitation energy refers to the transition between the two excited singlets.}
\label{fig:ZBW_ImChi}
\end{figure*}
The ZBW approximation amounts to the following simplification of the original Hamiltonian in Eqs.~(\ref{am1}-\ref{am3}):
\begin{align}
H_{\rm LR}&\approx 
-\sum_{\alpha}\left( \Delta_\alpha c^\dagger_{\alpha\uparrow}c^\dagger_{\alpha\downarrow}+\Delta^*_\alpha c_{\alpha\downarrow} c_{\alpha\uparrow}\right),\\
H_T&\approx\sum_{\alpha \sigma}(t_\alpha c^\dagger_{\alpha\sigma}d_\sigma+t^*_\alpha d^\dagger_\sigma c_{\alpha\sigma}),
\end{align}
by which all momentum sums are removed. Within this approximation, the Hilbert space of the two-channel problem with no restrictions on the dot-charge is only 64-dimensional and the Hamiltonian is readily diagonalized numerically. The Hamiltonian preserves electron parity, and all eigenstates can be conveniently sorted into states of even, or odd parity. Knowing the exact eigenenergies, $E_{n}$, and corresponding eigenstates, $|n\rangle$, the retarded current-current response function can be calculated using the Lehmann representation
\begin{align}\label{eq:zbwchiR}
\chi^R(\omega)=\sum_{nn'}\frac{e^{-E_n/T}}{Z}\frac{2\omega_{nn'}|\langle n'|\hat{I}|n\rangle|^2}{(\omega+i\eta)^2-\omega_{nn'}^2},
\end{align}
where $Z$ denotes the partition sum, $T$ the temperature (in units where $k_{B}=1$), $\omega_{nn'}=E_{n'}-E_{n}$, and $\eta$ is a positive infinitesimal.
Within the ZBW approximation, $\chi^R(\omega)$ comprises a sum of resonant terms, most of which should rightfully be ascribed to the continuum, which the ZBW approximation incorrectly describes as a discrete spectrum. In the zero-temperature limit, and restricting our attention to the transition from the ground state to the first excited state of same parity, the expression for the imaginary part ${\chi^R}''(\omega)=\delta \chi''(\omega)$ becomes particularly simple:
\begin{align}\label{eq:zbwimdchi}
\delta\chi_{\text{res}}''(\omega)=\pi|I^{\rm zbw}_{01}|^2\big(\delta(\omega+\omega_{01})-\delta(\omega-\omega_{01})\big).    
\end{align}
This is again the resonant contribution to the current response introduced previously, but now with current matrix element $|I^{\rm zbw}_{01}|^2=|\langle 1|\hat{I}|0\rangle|^2$ calculated within the ZBW approximation. The amplitude of $\delta\chi''_{\text{res}}$ at resonance ($\omega=\omega_{01}$) is plotted in Fig.~\ref{fig:ZBW_ImChi} for two different sets of parameters, corresponding to the cases $U=\Delta$ and $U=2\Delta$, respectively. For these parameters, a region of doublet ground state is found near $\varphi=\pi$, and for direct comparison with the corresponding plots in Figs.~\ref{fig:infgaptheta} and~\ref{fig:YSRY3}, we show only the matrix element for the singlet ground state.
Figs.~\ref{fig:ZBW_ImChi}(a-b) resemble the infinite-gap results in Fig.~\ref{fig:infgaptheta}, in which the matrix element remains finite even when decoupling the right lead ($\theta=0$). Figs.~\ref{fig:ZBW_ImChi}(c-d), on the other hand, have $U=2\Delta$, which is already large enough to suppress these terms and resemble the YSR scenario described above within the polarized-spin approximation (cf. Fig.~\ref{fig:YSRY3}).
\begin{figure}[t]
\includegraphics[width=\linewidth]{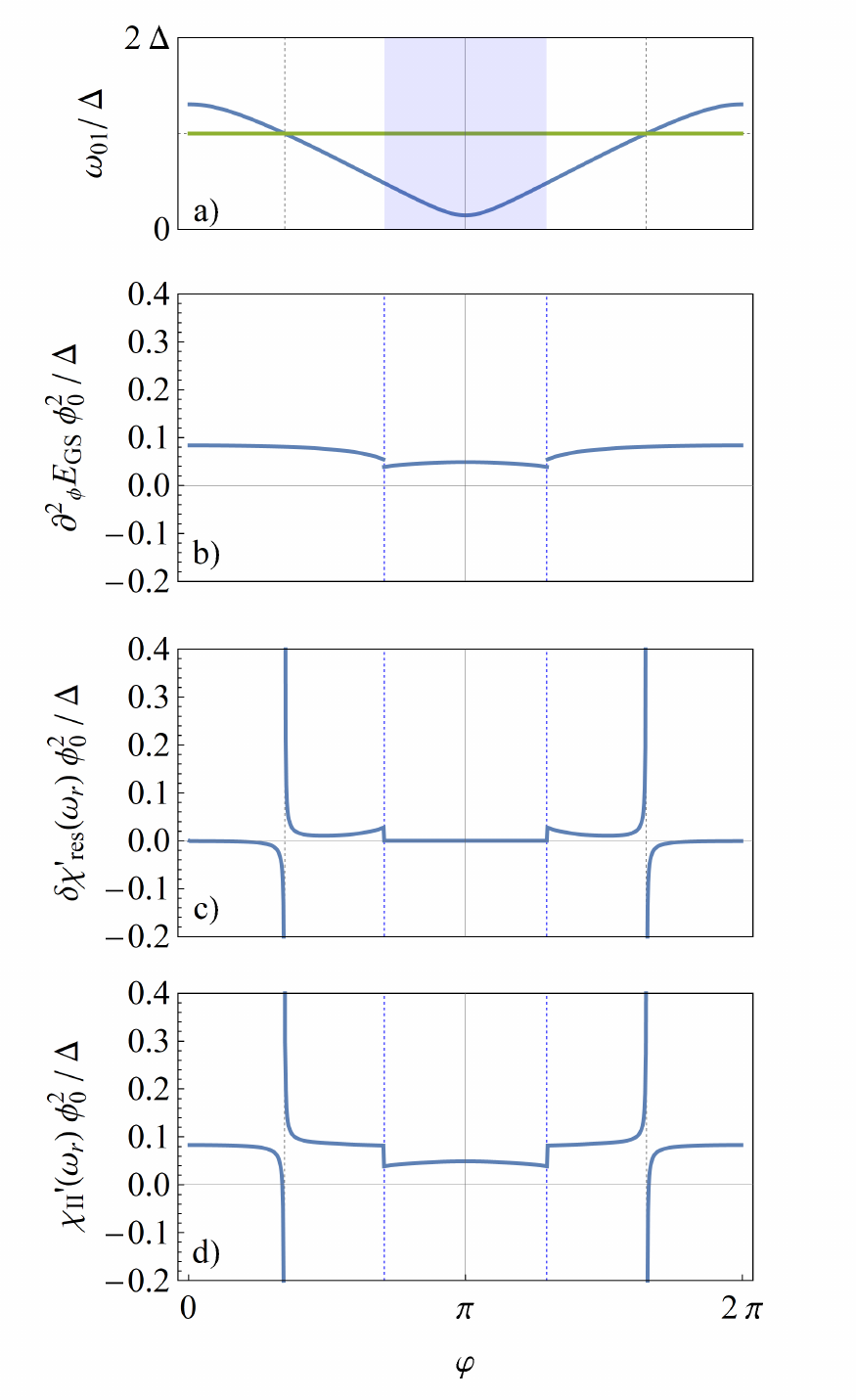}
\caption{(a) Singlet excitation energy corresponding to a transition between the two screened states. The shaded region and blue vertical dashed lines delimit the region around $\varphi=\pi$ in which the ground state is a doublet. (b) Inductive, and (c) resonant contributions to the total real part of the response function (d). Parameters are $\omega_{r}=\Delta, t=\Delta, U=2\Delta, x=0.25,$ and $\theta=\pi/4$, corresponding to the blue dashed line in Fig.~\ref{fig:ZBW_ImChi}(c). All curves are calculated within the ZBW approximation.}
\label{fig:ZBW_ReChi_YSR_2}
\end{figure}
\begin{figure}[t]
\includegraphics[width=\linewidth]{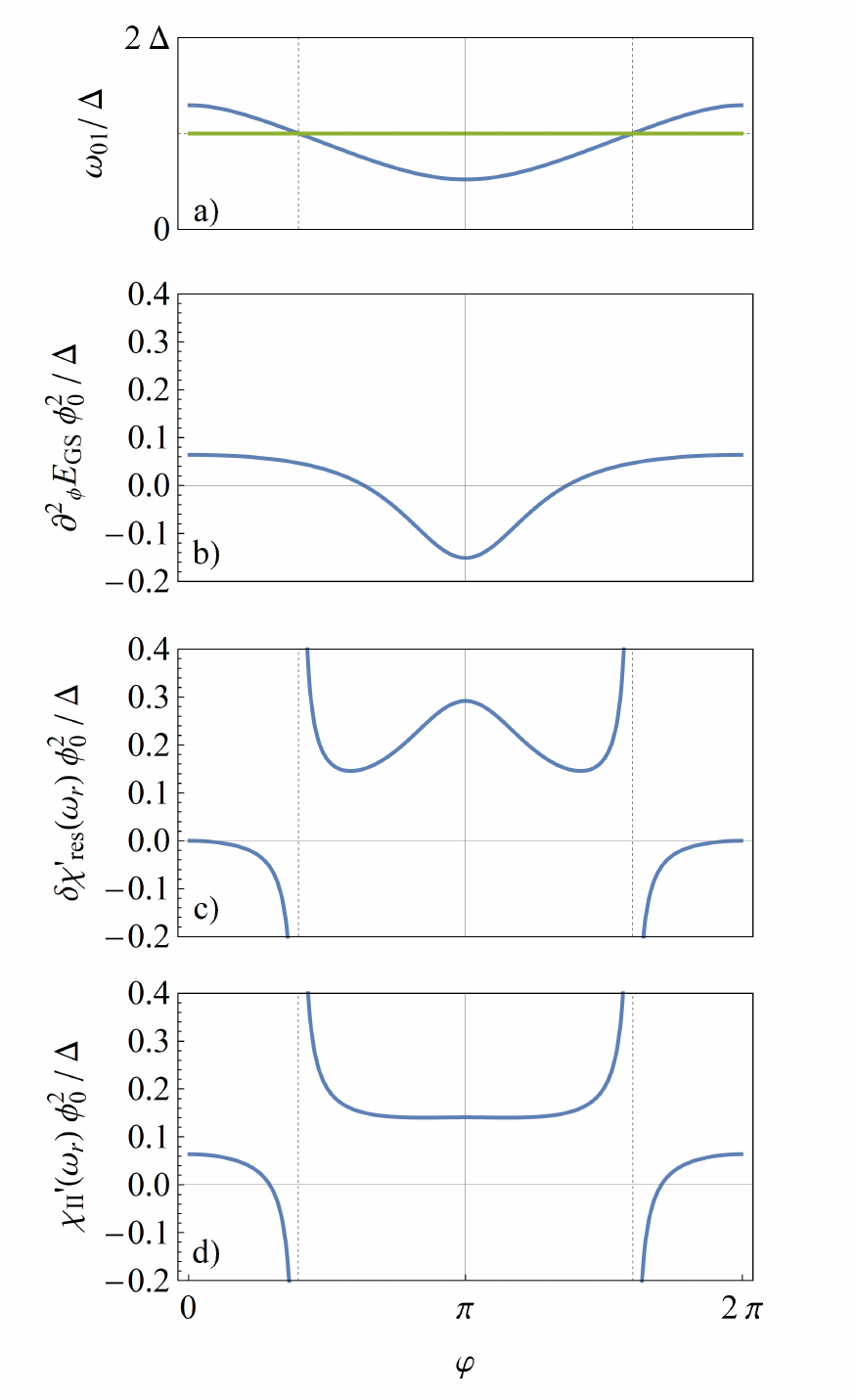}
\caption{(a) Singlet excitation energy corresponding to a transition between the two screened states. The ground state remains a singlet state for all values of $\varphi.$ (b) Inductive, and (c) resonant contributions to the total real part of the response function (d). Parameters are $\omega_{r}=\Delta, t=\Delta, U=2\Delta, x=0,$ and $\theta=\pi/6$, corresponding to the blue dashed line in Fig.~\ref{fig:ZBW_ImChi}(d). All curves are calculated within the ZBW approximation.}
\label{fig:ZBW_ReChi_YSR_1}
\end{figure}

The corresponding real part reads (assuming $\omega>0$) 
\begin{equation}\label{eq:zbwredchi}
\delta\chi_{\text{res}}'(\omega>0)=\frac{2|I^{\rm zbw}_{01}|^{2}}{\omega_{01}}\frac{\omega^2}{\omega^2-\omega_{01}^2},
\end{equation}
which is plotted in Figs. ~\ref{fig:ZBW_ReChi_YSR_2} and ~\ref{fig:ZBW_ReChi_YSR_1} alongside with the inductive contribution obtained numerically as $\partial^2_\phi E_{\mathrm{GS}}$ for two different sets of parameters, corresponding to the blue dashed lines in Figs.~\ref{fig:ZBW_ImChi}(c-d). As in the infinite-gap limit and polarized-spin approximation, these plots again show the characteristic avoided crossing at the resonance frequency. The continuum contribution is not accessible within the ZBW approximation as the continuum states remain discrete. 

Concerning the general features of the phase dependence of $\chi_{II}^{\prime}(\omega)$, the ZBW approximation interpolates between the proximitized limit ($U\ll\Delta$) captured by the infinite-gap approximation and the strongly interacting, Coulomb-blockaded, limit ($U\gg\Delta$) captured by the polarized-spin approximation. This remains a qualitative comparison insofar as the quantitative validity of the ZBW approximation is generally difficult to assess. Overall, however, comparison to numerical renormalization group (NRG) calculations~\cite{Steffensen2017, Grove-Rasmussen2018Jun, EstradaSaldana2018Dec}, has shown that the gate-dependent low-lying sub-gap states of the superconducting version of the Anderson model in Eqs.~(\ref{am1}-\ref{am3}) are captured very well by the ZBW approximation, upon adjusting the tunnelling amplitudes $t_{L/R}$ to accommodate for the missing density of states, $\nu_F$. As mentioned above, the ZBW approximation lacks a good description of the continuum states and therefore only captures the resonant contribution. 

\subsection{Nonequilibrium response}~\label{sec:Noneq}

Up to this point, we have only considered the linear microwave response of the system in its ground state. As demonstrated, this tends to enforce a large resonance frequency close to the gap, which is often difficult to attain experimentally. Actual experiments, however, have typically shown a parity flip time of the order of 20-200 $\mu$s~\cite{Janvier2015Sep, Hays2018Jul, Hays2020Nov, Hays2021Jul},
and with a measuring time, which is much smaller than this, the system will be measured in a given parity state, and the many points comprising a curve of dispersive shift vs. phase-difference, say, will typically sample the response from states of both parities. Whereas the non-equilibrium parity dynamics poses a device-dependent problem, which is interesting in its own right~\cite{Houzet2019Sep}, we shall restrict the present discussion to a few observations regarding the typical response from excited states. This can readily be done using the ZBW approximation, simply by replacing the thermal Boltzmann factors in Eq.~\ref{eq:zbwchiR} by non-equilibrium occupation numbers, or by considering the response term by term and studying instead the response of the individual many-body eigenstates. 

\begin{figure}
\centering
\includegraphics[]{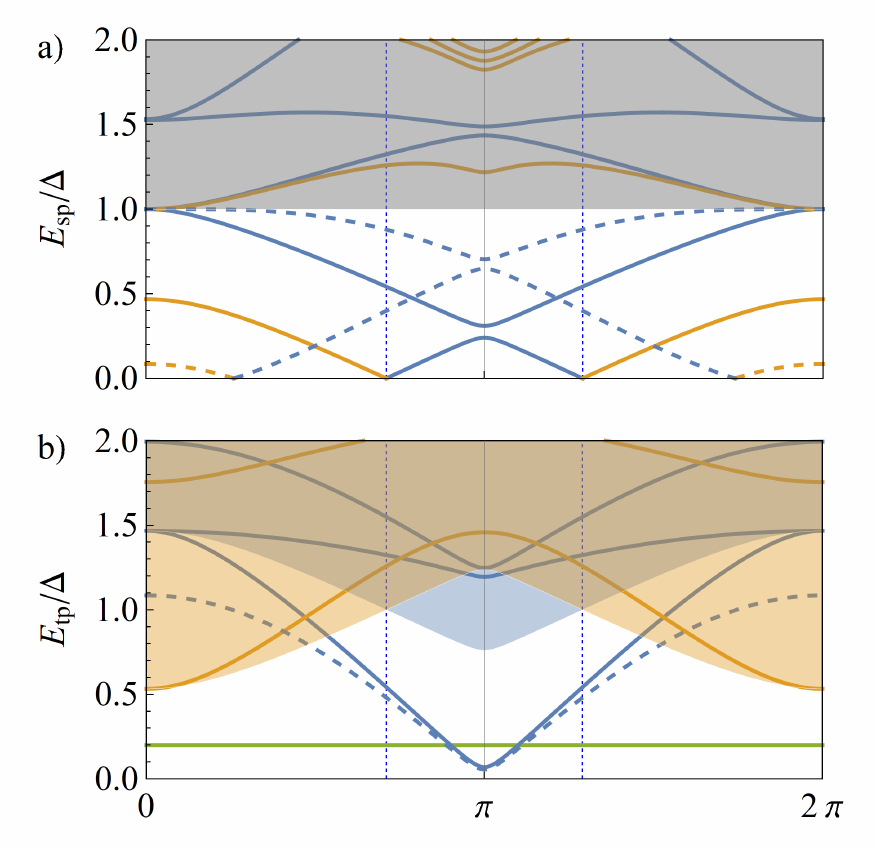}
\caption{(a) Single-particle excitation energies of parity-flipping transitions from the lowest-lying even(odd)-parity state to the odd(even) parity states, shown in orange(blue). Vertical blue dashed lines delimit the region of doublet ground state around $\varphi=\pi$. (b) Parity, and spin-conserving singlet (blue), and doublet (orange) two-particle excitation energies. Parameters are $x=0.1$, $\theta=\pi/4.1$. Full(dashed) curves are calculated within the ZBW (polarized-spin) approximation with $t=1.7\Delta$, and $U=8\Delta$ ($\Gamma/U=0.27$). Green line in (b) indicates the resonator frequency, $\omega_{r}=0.2\Delta$, used in Fig~\ref{fig:noneq2}.}
\label{fig:noneq1}
\end{figure}

\begin{figure}
\centering
\includegraphics[]{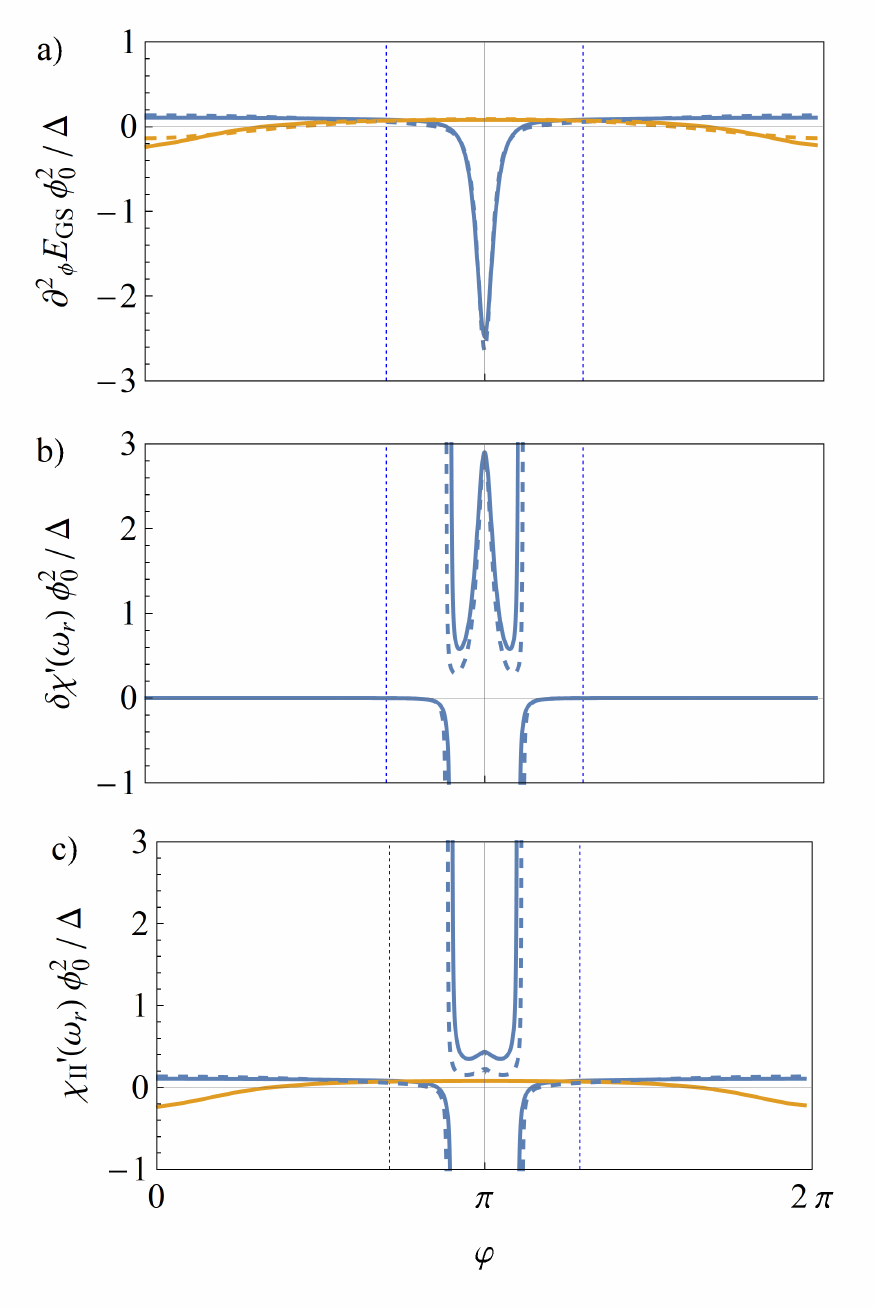}
\caption{ (a) Inductive contributions of the lowest-lying singlet (blue) and the doublet (orange) state, and (b) the corresponding resonant contributions to the total real part of the response function shown in (c). Parameters are the same as in Fig.\ref{fig:noneq1}}
\label{fig:noneq2}
\end{figure}

As was the case within the infinite-gap approximation, the ZBW approximation provides the exact eigenstates of the fully interacting problem, albeit in a reduced Hilbert space. Sorting these according to parity, i.e. electron number modulo two, we can infer the parity-flipping excitation spectrum, shown in Fig.~\ref{fig:noneq1}(a), as the lowest eigenenergy in the odd (even) sector subtracted from the eigenenergies of the even (odd) sector, shown here in blue (orange).
Parameters were chosen such that the ground state is, respectively, an even-parity spin-singlet state near $\varphi=0$ and $2\pi$, and an odd-parity spin-doublet state near $\varphi=\pi$, with the transition points indicated by vertical blue dashed lines. Likewise, Fig.~\ref{fig:noneq1}(b) shows the eigenenergies relative to the lowest eigenenergy within the even (blue) or odd (orange) parity sectors. These are the two-particle excitation energies of the parity, and spin-conserving transitions induced by the current operator and relevant to the microwave response. The blue shaded regions indicate the continua, which would emerge when including more states from the BCS leads. For the parity-flipping single-particle excitations, the continuum simply starts at $\Delta$, whereas the parity-conserving two-particle continuum involves a single-particle sub-gap excitation together with a single quasiparticle in the BCS leads.
Consequently, the onset of the two-particle continuum is at $\Delta+E_{\text{sp}},$ where $E_{\text{sp}}$ is the lowest-lying single-particle excitation energy, which is positive (negative) in the ground (excited) state. This leads to two different continuum thresholds for the even, and odd sector as seen in Fig.~\ref{fig:noneq1}(b). Additionally, the plot shows that the lowest doublet excitation energy is always above the continuum threshold, so that only the singlet transition is a resonant (sub-gap) excitation. These observations are analogous with the results obtained within the polarized-spin approximation, and correspond to the distinction made in Fig.~\ref{fig:admittance} between excitations at $\omega_{01}$ and $\omega_{02}$, respectively.

We note, that the singlet, and doublet two-particle excitations referred to in this work, are short for singlet-to-singlet, and doublet-to-doublet, as dictated by spin-conservation. These two types of parity-conserving transitions are often referred to as respectively pair-transitions and single-particle transitions~\cite{Tosi2019Jan, Metzger2021Jan}, being induced by respectively two- or single-quasiparticle terms in the current operator. 



Figs.~\ref{fig:noneq2}(a,b) again show the inductive, and resonant parts of the response function together with their sum in Fig.~\ref{fig:noneq2}(c). The response of the doublet (singlet) is shown in orange (blue). As only threshold doublet excitations are possible, the doublet state only provides an inductive response, and we note that the inductive response of the doublet states is typically smaller and flatter than that of the singlet states near $\varphi=\pi$, with curvature vanishing with $t/U$. 




In Figs.~\ref{fig:noneq1} and~\ref{fig:noneq2}, we have chosen a large charging energy, $U=8\Delta$, and a coupling ($t=1.7$) and particle-hole asymmetry ($x=0.1$) which are small enough to tentatively justify the Schrieffer-Wolff transformation and allow for a comparison to the polarized-spin approximation. To this end, Figs.~\ref{fig:noneq1} and~\ref{fig:noneq2} show not only ZBW results (solid lines), but also a set of dashed lines, which are calculated within the polarized-spin approximation, using the same parameters and tuning $\Gamma/U$ to 0.27, so as to ensure that both sub-gap two-particle excitation energies are in resonance with $\omega=0.2\Delta$, at the same value of $\varphi$. For the dashed lines, blue (orange) curves now refer to {\it screened (unscreened)} rather than {\it singlet (doublet)}. In contrast to the two-particle, the single-particle excitation energies do not match very well. This reflects a shortcoming of the polarized-spin approximation, which, unlike the infinite-gap, and the ZBW approximations at finite $U$, exhibits non-interacting quasiparticles. It is therefore only within the polarized-spin approximation that the two-particle excitation spectrum can be constructed from the single-particle spectrum.

In Figs.~\ref{fig:noneq2}(a-c) the response is dominated by the low-lying singlet transition near $\varphi=\pi$, deep inside the region of doublet ground state. As can be seen by comparing Figs.~\ref{fig:ZBW_ReChi_YSR_2} and \ref{fig:noneq2}, the main effect of including excited states in the calculation is the presence of the inductive, and resonant singlet responses close to $\varphi=\pi$ (inside the doublet region), where the singlet excitation energy is generally small, and the response is larger. From Figs.~\ref{fig:noneq2}(a-c), we also note that the moderate response of the excited singlet near $\varphi=\pi$ relies on a cancellation of much larger inductive, and resonant responses.

\section{Conclusion}~\label{sec:conclusion}

We have calculated the linear microwave response of superconducting cotunnel junctions based on Coulomb blockaded quantum dots, or magnetic impurities in a superconducting tunnel junction, modelled by a single-orbital Anderson model with two superconducting leads described by BCS mean-field theory. The calculations have been carried out using three different approximations. Firstly, the interacting infinite-gap limit was solved analytically, capturing the response due to dynamical proximity effect for $\Delta\gg U,\Gamma$. The current response was found to be consistent with the results of Ref.~\onlinecite{Kurilovich2021May}, and to vanish altogether for $U>2E_{A}$, when the ground state changes from even-parity spin-singlet to odd-parity spin doublet. Secondly, a Schrieffer-Wolff transformation was employed to obtain an effective low-energy exchange-cotunneling model, valid for $\Gamma,\Delta\ll U$ and $|x|\ll 1$. This effective model was studied within the polarized-spin approximation, building on the phase, and gate-dependent two-channel YSR bound states analyzed earlier in Ref.~\onlinecite{Kirsanskas2015Dec}. Analytical results for the inductive, and the resonant parts of the response function were obtained, while the contribution from the continuum was found numerically. Thirdly, we have employed the ZBW approximation to the full Anderson model, and demonstrated qualitatively that this approach interpolates between the results obtained within the infinite-gap, and the polarized spin approximations. 

Outside of the weak coupling limit, $\Gamma\ll U$, the strict validity of the ZBW approximation is generally hard to assess~\cite{Kirsanskas2015Dec}. Nevertheless, earlier comparisons~\cite{Grove-Rasmussen2018Jun, EstradaSaldana2018Dec} have shown excellent agreement in the sub-gap spectrum to results of numerically exact, and more demanding, NRG calculations, when allowing for a rescaling of $\Gamma$, which is often hard to assess directly by experiment anyway. It would be interesting to perform a detailed benchmarking against NRG, comparing current matrix elements, which rely on the details of the eigenstates. This may help consolidate the ZBW approach as a fast, versatile and reliable tool to analyze also microwave response functions for interacting S-QD-S junctions, and issue due warnings about potentially troublesome parameter regimes where the ZBW approximation fails. For example, the ZBW approximation does not capture the continuum contribution to the total response, which may be substantial for resonator frequencies which are not far below the continuum, as we have demonstrated within the polarized-spin approximation.

We have illustrated the results in a set of exemplary plots of the inductive, as well as the resonant contributions to the total response function. Within the polarized-spin approximation, we refer to screened and unscreened states, rather than spin-singlet and doublet states, and our comparison to the ZBW calculations substantiates the translation from one to the other. 
Only sub-gap excitations are accessible within the ZBW approximation, and we obtained a resonant contribution to the response function from the singlet state, as well as an inductive contribution from both singlet and doublet states. The two-particle singlet transition has its minimum excitation energy near $\varphi=\pi$, where it is no longer the ground state unless $T_{K}\gg\Delta$. This implies that an S-QD-S junction found in its ground state, will most likely only give rise to resonant frequency shifts from the singlet transition at very high resonator frequencies close to the gap. A junction with a finite parity decay time, on the other hand, may easily display a resonant shift corresponding to the singlet excitation energy, which is generally much smaller than the gap near $\varphi=\pi$, and vanishes altogether at $\varphi=\pi$ for a symmetrically coupled junction right at the particle-hole symmetric point. This transition is available even at weak coupling, where the junction remains a $\boldsymbol{\pi}$, corresponding to the sub-gap spectrum in Fig.~\ref{fig:Eysr}(a). A likely experimental outcome, for a junction sampling both parities, would therefore be a very weak non-resonant ground state response of high intensity superimposed by a strong phase-dependent low-frequency response of lower intensity. The intensity of the response from the different states, i.e. the number of points being measured with the system in that given state, will be decided by their relative parity lifetimes. The full nonequilibrium problem including the parity dynamics poses an interesting problem to pursue further in this regime of interacting QDs, due to either quasiparticle poisoning and relaxation, or for two-tone spectroscopy, where for example the gate voltage is oscillated~\cite{Tosi2019Jan, Hays2020Nov, Metzger2021Jan}. 

In recent years, superconducting junctions comprised of InAs wires with epitaxial Al have been studied intensively, both via DC bias spectroscopy or switching currents~\cite{Chang2015Mar, Grove-Rasmussen2018Jun, EstradaSaldana2018Dec, Saldana2020Nov, Kringhoj2020Jun}, and cQED techniques~\cite{Tosi2019Jan, Hays2018Jul, Hays2020Nov, Hays2021Jul, Kringhoj2020Jun}. In order to decide whether a given gated device behaves like a multi-channel wire (SNS) with Andreev bound states or more like a Coulomb blockaded QD with YSR states, with each their dependence on phase difference and gate voltage, it should be very illuminating to study devices with both DC, and AC probes available, together with both phase, and gate control, along the lines of the gatemon device studied in Ref.~\onlinecite{Kringhoj2020Jun}.


In closing, we note that two very recent cQED experiments on nanowire junctions both indicate the importance of interaction~\cite{Fatemi2021Dec, Canadas2021Dec}. Whereas the device studied in Ref.~\onlinecite{Canadas2021Dec} appears to be in a more open multi-channel regime with a small charging energy estimated to be $U/\Delta\sim 0.1$, the device studied in Ref.~\onlinecite{Fatemi2021Dec} is interpreted in terms of a larger charging energy, $U/\Delta\sim 0.75-3$. It should be interesting to try to fit the data in Ref.~\onlinecite{Fatemi2021Dec} within our strong-interaction modelling using the ZBW approximation.




\section{Acknowledgments} 
We acknowledge useful discussions with Valla Fatemi, and Pavel Kurilovich. The Center for Quantum Devices is funded by the Danish National Research Foundation. This work is supported by Novo Nordisk Foundation grant NNF20OC0060019 (CH). ALY acknowledges support from the Spanish AEI through grants PID2020-117671GB-I00 and through the “Mar\'ia de Maeztu” Programme for Units of Excellence in R\&D (Grant No. MDM-2014-0377) and by EU through grant no. 828948 (AndQC).

\appendix

\onecolumngrid
\section{YSR current matrix element}~\label{Appf}

The integrand in Eq.~\eqref{YSR_ReY} takes the form 
\begin{align}
\Bigg(&\frac{g_\sigma(\varepsilon, \sqrt{\Delta^2-(\varepsilon+i\eta)^2})}{D_\sigma(\varepsilon+i\eta)}-\frac{g_\sigma(\varepsilon, \sqrt{\Delta^2-(\varepsilon-i\eta)^2})}{D_\sigma(\varepsilon-i\eta)} \Bigg)\times\left(\epsilon\leftrightarrow \epsilon-\omega\right),
\end{align}
where $\eta$ denotes a positive infinitesimal and $D_\sigma(\varepsilon\pm i\eta)$ is the denominator of the total $G^{R/A}(\omega)$ in the combined Nambu and lead space. The function $g_\sigma$ in the numerator has a lengthy closed-form expression, which can be expanded to leading order in $\eta$ using that
\begin{align}
\sqrt{\Delta^2-(\varepsilon\pm i\eta)^2}=&\,\theta(\Delta-|\varepsilon|)\sqrt{\Delta^2-\varepsilon^2}\mp i \theta(|\varepsilon|-\Delta)\text{sgn}(\varepsilon)\sqrt{\varepsilon^2-\Delta^2},
\end{align}
which in turn allows the integrand to be rewritten as
\begin{align}
-4\Bigg(&\text{Re}[g_\sigma(\varepsilon)]\text{Im}\left[\frac{1}{D_\sigma(\varepsilon+i\eta)}\right]+\text{Im}[g_\sigma(\varepsilon)]\text{Re}\left[\frac{1}{D_\sigma(\varepsilon+i\eta)}\right]\Bigg)\times \left(\epsilon\leftrightarrow \epsilon-\omega\right).\label{eq:ImChiConts}
\end{align}
In the first term, the imaginary part can be taken by expanding the denominator to first order in $\eta$ as
\begin{align}
\Im\left[ \frac{1}{D_\sigma(\varepsilon+i\eta)}\right]&\approx
\Im\left[ \frac{1}{D_\sigma(\varepsilon)+i\eta D^\prime_\sigma(\varepsilon)}\right]=-\frac{\pi}{D^\prime_\sigma(\varepsilon)}\sum_{i=\pm}\delta(\varepsilon-E_{i\sigma}),
\end{align}
which is valid for $\varepsilon<\Delta.$ Eq.~\eqref{eq:ImChiConts} contains four terms. For $\epsilon, \omega-\epsilon<\Delta$ the only nonzero term is 
\begin{align}
-4\pi^2&\sum_{i,j=\pm}\frac{\text{Re}[g_\sigma(\varepsilon)]\text{Re}[g_\sigma(\varepsilon-\omega)]}{D^\prime_\sigma(\varepsilon)D^\prime_\sigma(\varepsilon-\omega)}\delta(\varepsilon-\omega-E_{j\sigma})\delta(\epsilon-E_{i\sigma}).
\end{align}
Inserting this in Eq.~\eqref{YSR_ReY} and introducing two step functions to ensure that $\epsilon, \omega-\epsilon<\Delta$ yields the following expression for the resonant contribution to $\delta\chi''(\omega),$ which only involves creation of quasiparticles in the subgap states
\begin{align}
\chi''_{\text{res}}(\omega)&=\pi\!\!\!
\sum_{\sigma; i,j=\pm}
\frac{\theta(E_{i\sigma})\theta(-E_{j\sigma})}
{D'_{\sigma}(E_{i\sigma})D'_{\sigma}(E_{j\sigma})}\int_{0}^{\omega}\!\!d\varepsilon f_\sigma(\varepsilon,\omega)
\delta(\omega-E_{i\sigma}+E_{j\sigma})
\delta(\varepsilon-E_{i\sigma})\label{YSRY3},
\end{align}
where we defined $f_\sigma(\varepsilon,\omega)=\text{Re}[g_\sigma(\varepsilon)]\text{Re}[g_\sigma(\varepsilon-\omega)].$ Evaluating the integral in Eq.~\eqref{YSRY3}, we obtain Eq.~\eqref{eq:ResContImChi}.

The function, $f_\sigma(\epsilon,\omega)$, takes the following form
\begin{align}
f_\sigma(\epsilon,\omega)=&\,\frac{e^{2}}{4}\sin^{2}(2\theta)
\Big[n(w,\epsilon,\omega)+n(-w,\epsilon,\omega)-m(w,\varphi,\epsilon,\omega)-m(-w,-\varphi,\epsilon,\omega)\nonumber\\
&-
\frac{1}{2}u\Delta^{2}\sin^{2}(2\theta)\Big(\epsilon(\epsilon-\omega)\psi^{2}+\xi(\epsilon)\xi(\epsilon-\omega)u^{2}\chi+u\left[\epsilon\,\xi(\epsilon-\omega)+(\epsilon-\omega)\xi(\epsilon)\right]\psi\cos(\varphi/2)\Big)\nonumber\\
&+2u\Delta^{2}{\rm Re}\left[\left(\xi(\epsilon-\omega)e^{i\varphi/2}+\sin^{2}(\theta)\left(2\sigma g(\epsilon-\omega)e^{i\varphi/2}+u\,\xi(\epsilon-\omega)\zeta(-\varphi)\right)\right)\right.\nonumber\\
&\left.\left.\hspace*{21mm}\times\left(\xi(\epsilon)e^{i\varphi/2}+\cos^{2}(\theta)\left(2\sigma g \epsilon e^{i\varphi/2} +u\,\xi(\epsilon)\zeta(\varphi)\right)\right)\right]\right],
\end{align}
written in terms of the following dimensionless quantities
\begin{align}
u&=w^2-g^2,\\
\chi&=1-\sin^2(2\theta)\sin^2(\varphi/2),\\
\psi&=e^{-i\varphi/2}(g+\sigma w)+e^{i\varphi/2}(g-\sigma w),\\
\zeta(\varphi)&=e^{i\varphi/2}\cos(\theta)^{2}+e^{-i\varphi/2}\sin(\theta)^{2},\\
\end{align}
and the functions
\begin{align}
\xi(\epsilon)&=\sqrt{\Delta^{2}-\epsilon^{2}},\\
l(w,\epsilon,\varphi)&=(g\sigma-w)\Delta^{2}e^{i\varphi}+(g\sigma+w)\epsilon^{2}+u\epsilon\,\xi(\epsilon),\\
m(w,\varphi,\epsilon,\omega)&=\frac{1}{4}\sin^{2}(2\theta)(g\sigma+w)^{2}l(w,\epsilon-\omega,\varphi)l(w,\epsilon,\varphi),\\
n(w,\epsilon,\omega)&=(g\sigma+w)^{2}\left[\epsilon\,\xi(\epsilon)(1+u\cos^{2}(\theta))-(g\sigma-w)\sin^{2}(\theta)\xi(\epsilon)^{2}+2\epsilon^{2}g\sigma\cos^{2}(\theta)\right]\nonumber\\
&\hspace*{5mm}\times\left[2g\sigma(\epsilon-\omega)^{2}\sin^{2}(\theta)-(g\sigma-w)\cos^{2}(\theta)\xi(\epsilon-\omega)^{2}+(\epsilon-\omega)\xi(\epsilon-\omega)\left(1+u\sin^{2}(\theta)\right)\right].
\end{align}

\twocolumngrid

\bibliographystyle{apsrev4-2}
\bibliography{paper_bib_admit}

\end{document}